\documentclass[superscriptaddress,showpacs,amssymb,10pt,reprint,aps,prd,longbibliography,nofootinbib,floatfix]{revtex4-2}

\usepackage{graphicx,epsfig,amssymb,times} 
\usepackage{amsmath,amsfonts}
\usepackage{bm}
\usepackage{epstopdf}
\usepackage{hyperref}
\usepackage[caption=false]{subfig}
\usepackage[usenames]{color}   
\usepackage[dvipsnames]{xcolor}
\usepackage[normalem]{ulem}

\DeclareMathOperator{\sech}{sech}
\definecolor{coolblack}{rgb}{0.0, 0.18, 0.39}
\definecolor{darkred}{rgb}{0.5,0,0}
\definecolor{darkgreen}{rgb}{0,0.5,0}
\definecolor{darkblue}{rgb}{0,0,0.5}
\definecolor{lapislazuli}{rgb}{0.15, 0.38, 0.61}
\definecolor{venetianred}{rgb}{0.78, 0.03, 0.08}
\definecolor{bleudefrance}{rgb}{0.19, 0.55, 0.91}
\definecolor{dogwoodrose}{rgb}{0.84, 0.09, 0.41}
\hypersetup{colorlinks=true, citecolor=darkblue, linkcolor=darkblue, 
urlcolor = darkblue}

\begin{document}

\title{\large Absorption spectrum and greybody factors of charged black holes in loop quantum gravity}
	
\author{Marco A. A. de Paula}
\email{marcodepaula@ufpa.br}
\affiliation{Faculdade de Ciências Naturais, Universidade Federal do Par\'a, Campus Universit\'ario do Tocantins-Cametá, 68400-000, Camet\'a, Par\'a, Brazil}
\affiliation{Departamento de F\'isica, Universidade Federal da Para\'iba, 58051-970, Jo\~ao Pessoa, Para\'iba, Brazil.}

\author{Valdir B. Bezerra}
\email{valdir@fisica.ufpb.br}
\affiliation{Departamento de F\'isica, Universidade Federal da Para\'iba, 58051-970, Jo\~ao Pessoa, Para\'iba, Brazil.}
    
\author{Luiz C. S. Leite}
\email{luiz.leite@ifpa.edu.br}
\affiliation{Campus Altamira, Instituto Federal do Par\'a, 68377-630, Altamira, Par\'a, Brazil.}

\begin{abstract}

In the last few decades, singularity-free black holes (BHs) obtained in the framework of Loop Quantum Gravity (LQG) have gained attention in the literature. These compact objects replace the classical singularity with a transition hypersurface called the bounce radius and stand out as potential scenarios for exploring the imprints of LQG in BH physics. Although scalar perturbations in the vicinity of LQG-based BHs are currently being studied, the absorption spectrum has not yet been analyzed in detail. In this work, we present an in-depth investigation of the absorption properties of massless test scalar fields by a charged LQG BH, aiming to better understand the role played by the quantum and charge parameters of the BH spacetime. Using a numerical approach, we compute the absorption cross section (ACS) of the massless scalar wave for arbitrary values of the frequency of the incident wave. We find that the behavior of the ACS as we increase the quantum parameter indicates that the peaks and troughs of the total ACS exhibit opposite behaviors, i.e., the curve related to the highest peak corresponds to the deepest troughs. Moreover, we show that the ACS decreases as we consider higher values of the BH charge-to-mass ratio. This is in stark contrast to the behavior of the absorption spectrum as we vary the quantum parameter. We also draw comparisons with the Reissner-Nordström (RN) BH, exploring the situations where LQG and RN BHs can have the same absorption properties. Furthermore, we find excellent agreement between our numerical results and the well-known classical and semiclassical approximations for the total ACS in their corresponding limits. For completeness, we also investigate the greybody factors. Our results can be viewed as a first step toward a better understanding of the absorption properties of LQG-inspired BHs.

\end{abstract}

\date{\today}

\maketitle

\section{Introduction}

General Relativity (GR) is a robust theory of gravity corroborated by observational data. Simply put, GR is a geometric theory of gravity in which gravity is understood as the curvature of spacetime due to the presence of matter and energy~\cite{kox1997collected}. Over the past hundred years, GR has passed several experimental tests successfully and has also led to the prediction of new astrophysical objects and phenomena~\cite{wald2010general}. In this respect, we can mention, for example, a collection of gravitational-wave detections~\cite{Collaboration2025GWTC40AI}. Other important recent experimental results associated with GR were obtained by the \textit{Event Horizon Telescope Consortium}. These include the first shadow image of a supermassive BH at the center of the Messier 87 galaxy~\cite{EventHorizonTelescope:2019dse} and, more recently, the shadow of a supermassive BH at the center of the Milky Way galaxy~\cite{EventHorizonTelescope:2022wkp}. 

These recent observational results supporting GR strongly indicate that BHs are part of our physical reality. These objects are simple in the sense that they are characterized by a one-way membrane, called the event horizon, and described by only three parameters in the electrovacuum (mass, charge, and angular momentum)~\cite{heusler1996black}. However, within GR, they have an intrinsic (or curvature) singularity at their core. This represents a major challenge to Einstein's theory, since the known laws of physics are no longer applicable in these pathologies. Moreover, this is not the only challenge to GR, as the theory also fails to properly explain, for example, the earlier stages of the universe~\cite{Hawking:1970zqf} and the last stages of BH evaporation~\cite{Hawking:1975vcx}.

Some of the drawbacks of GR, such as those mentioned above, are widely believed to be limitations of its classical formulation. Therefore, it is expected that these problems will be solved in a full quantum gravity theory. Although we have not yet found such a formulation, it is possible to modify the mathematical machinery of Einstein's theory to incorporate quantum mechanics effects and circumvent some of the classical problems of GR. Among such modifications, the Loop Quantum Gravity (LQG) theory stands out as one of the most promising. Roughly speaking, this theory consists of a non-perturbative quantization of GR using the Ashtekar-Barbero variables and is built on the space of holonomies~\cite{Ashtekar:2004eh,Thiemann:2007pyv,gambini2011first,Rovelli:2014ssa}.

BH geometries can be interesting scenarios for exploring the LQG theory and revealing some of its features. Nevertheless, finding BH solutions within the scope of LQG is a daunting effort. Currently, several schemes for finding LQG-inspired BHs have been proposed; see, for example, Refs.~\cite{Corichi:2015xia,Han:2022rsx,Zhang:2023lqb,Perez:2017bhl,Zhu:2024wic,AlonsoBardaji:2021qsb,AlonsoBardaji:2022nsh} and references therein. The majority of the schemes share a common characteristic: they introduce a transition hypersurface within the BH. At this hypersurface, the spacetime tunnels from a BH to a white hole. In other words, the classical singularity is replaced with a transition surface that connects a trapped region and an anti-trapped region. This transition surface is often called the bounce radius. 

In this work, we are interested in the charged LQG BH derived in Ref.~\cite{Borges:2023fub}, which applies the uniparametric polymerization scheme~\cite{AlonsoBardaji:2021qsb,AlonsoBardaji:2022nsh}. In this scheme, the holonomy corrections are introduced through a canonical transformation, and the corresponding Hamiltonian is regularized (see also Ref.~\cite{Sobrinho:2022zrp} for details). This makes the derivation of singularity-free BH solutions, with an interior region that contains a spacelike hypersurface replacing the classical singularity, feasible. The hypersurface effectively acts as a bounce surface, governed by a unique quantum parameter that connects a white hole to a BH, and the corresponding exterior geometry is asymptotically flat. Moreover, in this approach, the angular sector of the line element maintains its classical form, implying that the horizon area is the same as the classical one. Besides that, the bounce radius is always within the BH.

The potential signatures of LQG in BH physics can be investigated by analyzing how LQG-based BHs interact with their surroundings. This is a fertile ground because BHs are expected to be surrounded by matter distributions in the astrophysical environment~\cite{Narayan:2005ie}. In this context, scalar perturbations have been studied in the background of LQG-inspired BHs from the perspective of quasinormal modes~\cite{Livine:2024bvo,Bolokhov:2023bwm,Moreira:2023cxy,Gingrich:2024tuf,Daghigh:2020fmw,Zhu:2024wic}, but the absorption spectrum of charged LQG BHs has not yet been investigated\footnote{Notice also that despite their simplicity, scalar fields are also relevant from the astrophysical point of view. They are important in the inflation era~\cite{Lidsey:1995np}, can be used to describe dark energy~\cite{Peebles:2002gy,Padmanabhan:2002ji}, and can be treated as possible alternative dark matter candidates (see Ref.~\cite{Barranco:2011eyw} and references therein).}. This can be accomplished by computing the total absorption cross section (ACS) ---  the ratio between the flux of the field that goes into the BH and the current of the incident planar wave~\cite{futterman1988scattering} --- which is closely connected with the greybody factors (GBFs)~\cite{Crispino:2013pya,Heidari:2024bvd}. The GBF measures the absorption probability of the BH. Since the late 1960s, the study of the absorption properties of spherically symmetric BHs has been done in several distinct scenarios (see, for example, Refs.~\cite{Unruh:1976fm,Jung:2004yh,Crispino:2009zza,Crispino:2009ki,Benone:2014qaa,Benone:2015bst,Delhom:2019btt,Lima:2020auu,Anacleto:2017kmg,Huang:2019ptr,Li:2021epb,Xavier:2021sje,Macedo:2015ikq,Anacleto:2019tdj,Anacleto:2022shk,Macedo:2014uga,sanchez2018scattering,Fernando:2016ksb,Paula:2020yfr,dePaula:2023muc,dePaula:2023xie,Magalhaes:2020pyp,Heidari:2024bkm} and references therein).  

Here, we present an in-depth investigation of the absorption spectrum of massless scalar waves in the background of the well-motivated static and spherically symmetric charged LQG BH derived in Ref.~\cite{Borges:2023fub}. We aim to understand the role played by the quantum parameter and the BH charge in the absorption properties of the charged LQG BH by combining analytical and numerical techniques. For completeness, we also investigate the GBFs. The remainder of this paper is organized as follows. In Sec.~\ref{sec:lqgbh}, we review the mathematical machinery used to derive the BH solution obtained in Ref.~\cite{Borges:2023fub} in LQG theory (\ref{subsec:pre}), and also explore some physical and geometrical properties of the BH solution (\ref{subsec:cbhg}). The classical and semiclassical approximations for the ACS are presented in Sec.~\ref{sec:ga}. In Sec.~\ref{sec:we}, we introduce the Klein-Gordon equation, where we analyze the scalar wave (\ref{subsec:sw}) and present the total ACS and GBFs (\ref{subsec:acsgbf}). A selection of our main results is discussed in Sec.~\ref{sec:mr}. Finally, in Sec.~\ref{sec:remarks}, we present our final remarks. Throughout this paper, we use natural units ($G = c = \hbar = 1$) and metric signature $+2$.

\section{Loop quantum gravity black holes}\label{sec:lqgbh}

In this section, starting from the classical Hamiltonian, we review the derivation of the charged LQG BH obtained in Ref.~\cite{Borges:2023fub}. We also discuss the main properties of this geometry, analyzing the metric function and curvature scalars.

\subsection{Preliminaries}\label{subsec:pre}

The classical Reissner-Nordström (RN) Hamiltonian, using the formalism introduced in Ref.~\cite{Ashtekar:2018cay}, can be written as~\cite{Borges:2023fub}
\begin{equation}
\label{Hcl}H_{\rm{cl}} = -\dfrac{1}{2 \gamma} \left[2c p_{c} +\left(b+\dfrac{\gamma^{2}}{b}-\dfrac{\gamma^{2}Q^{2}}{bp_{c}} \right)p_{b}\right].
\end{equation}
The parameters $b$ and $c$ are related to the full Ashtekar-Barbero connection, and are called connection variables, while the parameters $p_{c}$ and $p_{b}$ are the so-called triad variables~\cite{Ashtekar:2018cay}. They are related to the Hamiltonian framework of LQG and are canonically conjugate to the connection variables $b$ and $c$, satisfying the following Poisson brackets: 
\begin{equation}
\{b,p_{b}\} = \gamma \quad \text{and} \quad \{c,p_{c}\} = 2 \gamma. 
\end{equation}
The quantity $\gamma$ is the Barbero-Immirzi parameter~\cite{Immirzi:1996dr,BarberoG:1994eia}, which arises from the transition from classical to quantum theory and, roughly speaking, measures the size of the quantum of area. Moreover, $Q$ is a parameter related to the BH charge.

The line element associated with this setup can be written in the following form
\begin{equation}
ds^{2} = -N^{2}d\tau^{2} + \dfrac{p_{b}^{2}}{L_{0}^{2}|p_{c}|}dx^{2}+|p_{c}|d \Omega^{2},
\end{equation}
where $L_{0}$ is the fiducial length, $d\Omega^{2} = d\theta^{2}+\sin^{2}\theta d\varphi^{2}$, and $\tau$ is a time coordinate. The parameter $N$ is given by
\begin{equation}
N = \dfrac{\gamma \sqrt{p_{c}}}{b}.    
\end{equation}

Following the polymerization approach, we can promote canonical variables to quantum operators by performing a suitable transformation~\cite{Gambini:2021uzf}. For our purposes, we promote only the variable $b$, keeping $c$ classical. Thus, we have~\cite{Borges:2023fub}
\begin{equation}
b \rightarrow \dfrac{\sin(\delta_b b)}{\delta_b} \quad \text{and} \quad p_{b} \rightarrow \dfrac{p_{b}}{\cos(\delta_b b)},
\end{equation}
where $\delta_b$ is the polymerization parameter, which incorporates quantum geometry corrections from LQG into the classical equations. The classical results are obtained in the limit $\delta_{b} \rightarrow 0$. Then, by introducing the regularization factor given by
\begin{equation}
\label{regufactor}\dfrac{\cos(\delta_b b)}{\sqrt{1+\gamma^{2}\delta_b^{2}}},
\end{equation}
we can construct an effective Hamiltonian, namely,
\begin{widetext}
\begin{equation}
H_{\rm{eff}} = -\dfrac{1}{2\gamma \sqrt{1+\gamma^{2}\delta_b^{2}}}\left[2cp_c \cos(\delta_b b)+\left(\dfrac{\sin(\delta_b b)}{\delta_b}+\dfrac{\gamma^{2}\delta_b}{\sin(\delta_b b)}-\dfrac{\gamma^{2}\delta_b Q^{2}}{\sin(\delta_b b)p_c}\right)p_b \right].
\end{equation}
\end{widetext}
The effective Hamiltonian is obtained from the linear combination of the classical Hamiltonian~\eqref{Hcl} and the regularization factor~\eqref{regufactor}. This factor is introduced because we need a covariant representation of the spacetime~\cite{Gambini:2021uzf}. In summary, this is the so-called uniparametric polymerization scheme~\cite{AlonsoBardaji:2021qsb,AlonsoBardaji:2022nsh,Borges:2023fub}. 

One can write Hamilton's equations in terms of the Poisson brackets~\cite{greiner2013quantum}, given by
\begin{equation}
\label{hamieq}\dot{x} = \{x,H_{\rm{eff}}\},
\end{equation}
where $x \in (b, c, p_b, p_c)$ and the dot denotes differentiation with respect to $\tau$. Moreover, following Ref.~\cite{goldstein2002classical}, for a physical system with two degrees of freedom, one can show that
\begin{align}
\nonumber \{f, g\} = & \left( \frac{\partial f}{\partial b} \frac{\partial g}{\partial p_b} - \frac{\partial f}{\partial p_b} \frac{\partial g}{\partial b} \right) \{b, p_b\} +\\
\label{genfg}&\left( \frac{\partial f}{\partial c} \frac{\partial g}{\partial p_c} - \frac{\partial f}{\partial p_c} \frac{\partial g}{\partial c} \right) \{c, p_c\},
\end{align}
where $f$ and $g$ are arbitrary functions. Notice that in our case, we have the pairs $(b, p_b)$ and $(c, p_c)$. Inserting Eq.~\eqref{genfg} into Eq.~\eqref{hamieq}, we find a set of four dynamical equations, given by
\begin{widetext}
\begin{align}
\label{de1} \dot{b} = \ & \{b,H_{\rm{eff}}\} = -\frac{1}{2\sqrt{1+\gamma^2\delta_b^2}} \left( \frac{\sin(\delta_b b)}{\delta_b} + \frac{\gamma^2 \delta_b}{\sin(\delta_b b)} - \frac{\gamma^2 \delta_b Q^2}{\sin(\delta_b b) p_c} \right), \\
\label{de2} \dot{c} = \ & \{c,H_{\rm{eff}}\} = -\frac{1}{\sqrt{1+\gamma^2\delta_b^2}} \left[ 2 c \cos(\delta_b b) + \frac{\gamma^2 \delta_b Q^2 p_b}{\sin(\delta_b b) p_c^2} \right], \\
\label{de3} \dot{p_b} = \ & \{p_b,H_{\rm{eff}}\} = \frac{1}{2\sqrt{1+\gamma^2\delta_b^2}} \left[ -2 c p_c \delta_b \sin(\delta_b b) +  \left( 1 - \frac{\gamma^2 \delta_b^2}{\sin^2(\delta_b b)} + \frac{\gamma^2 \delta_b^2 Q^2}{p_c \sin^2(\delta_b b)} \right)p_b \cos(\delta_b b) \right], \\
\label{de4} \dot{p_c} = \ & \{p_c,H_{\rm{eff}}\} = \frac{2 p_c \cos(\delta_b b)}{\sqrt{1+\gamma^2\delta_b^2}}.
\end{align}
\end{widetext}

We can combine Eqs.~\eqref{de1} and~\eqref{de4}, eliminating the dependence on the parameter $\tau$. Then, one can show that
\begin{equation}
\label{sol1}\frac{\sin^2(\delta_b b)}{\gamma^2 \delta_b^2} = \frac{2M}{\sqrt{p_c}} - 1 - \frac{Q^2}{p_c}
\end{equation}
is a solution of the resulting first-order differential equation, with $M$ being the BH mass. By inserting Eq.~\eqref{sol1} into Eq.~\eqref{de4} and solving the differential equation, we find that
\begin{align}
\nonumber p_{c} = \ & \frac{e^{-2\tau}}{4b_0^4(b_0+1)^2M^{2}} \{ (b_0+1)M^2[b_0-1+(b_0+1)e^\tau]^2\\
\label{sol2} & - (b_0-1)b_0^2 Q^2 \}^{2},
\end{align}
where we fixed the integration constant in order for $Q = 0$ to lead to the uncharged solution~\cite{Sobrinho:2022zrp} and defined
\begin{equation}
\label{b0def}b_{0} = \sqrt{1+\gamma^{2}\delta_b^{2}}.
\end{equation}

\begin{figure}[!htbp]
\begin{centering}
    \includegraphics[width=1\columnwidth]{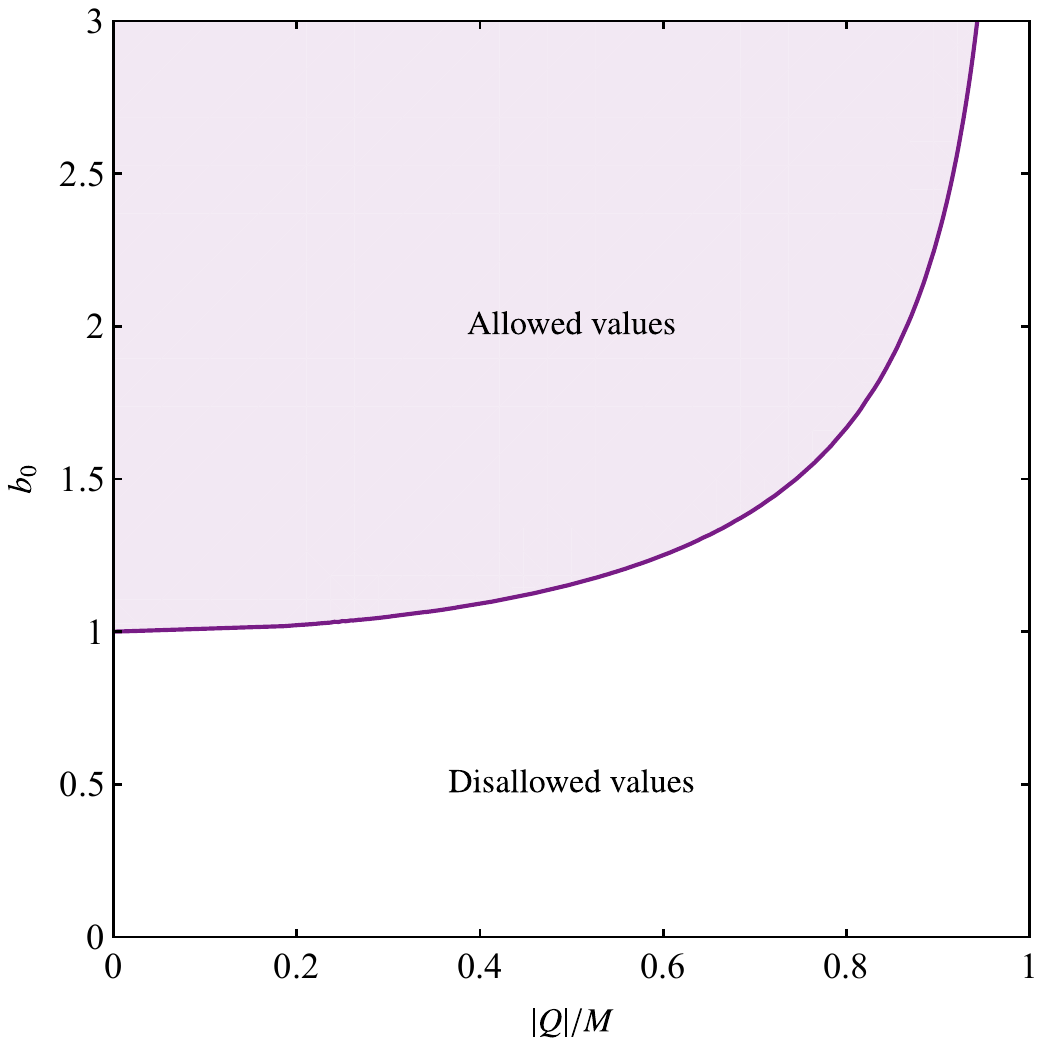}
    \caption{Allowed and disallowed values of $|Q|/M$ and $b_0$. The purple curve denotes the boundary where the relation given by Eq.~\eqref{constraintQ} is saturated. The highest value allowed by $|Q|/M$, for $b_{0}$ up to $3$, is $2\sqrt{2}/3$.}
    \label{av}
\end{centering}
\end{figure}

In LQG, the corrections introduced by quantum theory provide a bounce radius, denoted by $r_{0}$, which corresponds to the minimum physical radius reached by the geometry within the BH before it begins to expand again. This replaces the classical central singularity with a ``regular quantum transition''. In other words, the bounce radius connects the BH to a white hole. The areal radius is defined as $r \equiv \sqrt{p_c}$, and the minimum area can be obtained from $\dot{p}_{c} = 0$, which leads to
\begin{align}
\nonumber r_{0} & \equiv \sqrt{p_{c,\rm{min}}} \\
\label{r0}& = \frac{\left(b_0^2-1\right)M}{b_0^2} \left(1 + \sqrt{1 - \dfrac{b_0^2 Q^2}{\left(b_0^2-1\right)M^{2}}} \right).
\end{align}
Notice that the classical case (Maxwell's theory) is given by $b_{0} = 1$. In this situation, $r_{0}$ tends to the standard central singularity. Moreover, since $r_{0} > 0$, we find that
\begin{equation}
\label{constraintQ}|Q| \leq M \dfrac{\sqrt{b_{0}^{2}-1}}{b_{0}}.
\end{equation}
Therefore, the LQG corrections introduce an additional constraint on the possible values of the BH charge-to-mass ratio. In Fig.~\ref{av}, we display the allowed and disallowed values of $|Q|/M$ and $b_0$ associated with the condition given by Eq.~\eqref{constraintQ}. For simplicity, in this work, the highest value of $b_0$ that we will consider is $b_0 = 3$.

\subsection{Charged BH geometries}\label{subsec:cbhg}

To find the corresponding line element associated with the charged LQG BH spacetime, we need to use the remaining dynamical equations, apply a suitable change of variables, and perform an analytic continuation. Notice that
\begin{equation}
-N^2 d\tau^2 = -\dfrac{\gamma^{2}\delta_{b}^{2}b_{0}^{2}}{\sin^2(\delta_b b)\cos^2(\delta_b b)}dr^2,
\end{equation}
where we use $d\tau = (2 \sqrt{p_c}/\dot{p}_c)dr$ and Eq.~\eqref{de4}. The trigonometric terms can be factored using Eq.~\eqref{sol1}, leading to
\begin{equation}
\label{grr}-N^2 d\tau^2 = \left[f(r)\left(\dfrac{1}{b_0^2}-\dfrac{(1-b_0^2)f(r)}{b_0^2}\right)\right]^{-1}dr^2,
\end{equation}
with $f(r)$ being the metric function of the RN geometry, i.e.,
\begin{equation}
f(r) = 1 - \frac{2M}{r} + \frac{Q^2}{r^2}.
\end{equation}
Nevertheless, it is instructive to write the component $g_{rr}$ of the metric tensor as a function of $r_{0}$. Thus, we find that
\begin{equation}
\label{grr2}-N^2 d\tau^2 = \left[f(r)\left( 1 - \dfrac{r_0}{M} g(r) \right)\right]^{-1}dr^2,
\end{equation}
where
\begin{equation}
g(r) = \dfrac{\left(1-f(r)\right)}{1 + \sqrt{1 - \dfrac{b_0^2 Q^2}{(b_0^2 - 1) M^2}}}.
\end{equation}

The necessary steps to obtain the appropriate expression for the component $g_{tt}$ of the metric tensor are lengthy. For simplicity, we outline the main steps. By using the effective Hamiltonian constraint $H_{\rm{eff}} = 0$, isolating the dynamical variables, and performing the appropriate analytical continuation as derived and detailed in Refs.~\cite{Borges:2023fub,Sobrinho:2022zrp}, the resulting line element for the charged LQG BH spacetime is given by
\begin{equation}
\label{LEfinal}ds^2 = - f(r) dt^2 + \dfrac{1}{f(r)\left( 1 - \dfrac{r_0}{M} g(r) \right)} dr^2 + r^2 d\Omega^2.
\end{equation}

The line element~\eqref{LEfinal} is asymptotically flat and shares the same horizons as the classical theory, i.e.,
\begin{equation}
\label{horizonss}r_{\pm} = M\pm\sqrt{M^{2}-Q^{2}},
\end{equation}
where $r_+$ and $r_{-}$ are the event and Cauchy horizons, respectively. Notice that from the classical point of view, the condition $Q = M$ still mathematically defines the classical situation where the Cauchy and event horizons coincide. However, for the charged LQG BH spacetime studied in Ref.~\cite{Borges:2023fub}, a BH cannot generally reach $Q = M$ due to quantum geometry effects [the bounce radius restricts the possible values of $Q/M$, see Eq.~\eqref{constraintQ}]. In order to obtain the correct maximally charged case in this context, we impose the minimal area condition given by~\cite{Modesto2010}
\begin{equation}
\label{areagap}4\pi r_{0}^{2} = 4\sqrt{3}\pi \gamma.
\end{equation}
In other words, we impose that the transition surface must have a minimum area corresponding to the LQG area gap. In Ref.~\cite{Borges:2023fub}, the authors set $\gamma = \sqrt{3}/6$. However, the overall conclusions do not strictly depend on this particular choice. Using Eq.~\eqref{r0}, it is possible to show that Eq.~\eqref{areagap} leads to
\begin{equation}
\label{deltab}\delta_{b}^{2} = \dfrac{12}{2\sqrt{2}M-2Q^{2}-1}.
\end{equation}

Combining Eqs.~\eqref{constraintQ} and~\eqref{deltab}, we find that\footnote{The reader may find the dimensionality of the expression below odd, but note that we are using natural units here, so the Planck length $l_P = \sqrt{\hbar G/c^{3}}$ is dimensionless. If we used only geometric units, we would obtain $Q^{2}\leq (\sqrt{2}M l_{P}/2)$. This particular choice of units masks that in LQG theory, the extremal charge limit may be scale-dependent, i.e., it can fundamentally depend on its absolute mass relative to the Planck scale.}
\begin{equation}
\label{ext}Q^{2}\leq \dfrac{\sqrt{2}}{2}M.
\end{equation}
Notice that Eq.~\eqref{deltab} also implies that
\begin{equation}
\label{constraintQM}2\sqrt{2}M > 2Q^{2}+1,
\end{equation}
otherwise, $\delta_b$ diverges or takes on complex values. If we now combine Eqs.~\eqref{horizonss} and~\eqref{constraintQM}, we find that
\begin{equation}
\label{r0constraint}r_{-} < r_{0} < r_{+}.
\end{equation}
Therefore, the bounce radius lies within the event horizon but outside the Cauchy horizon. In other words, an observer crossing the event horizon can never hit the Cauchy horizon at $r_{-}$ because the spacetime itself bottoms out at $r_0$ and bounces. Notice that for the charged LQG BH spacetime, if we impose the classical charge condition for an extremely charged BH, i.e., $Q = M$, by using Eq.~\eqref{constraintQM}, we find that $Q  = \sqrt{2}/2$. In other words, this is the only configuration where the quantum bounds allow $f(r) = 0$ and $f^{\prime}(r) = 0$, where the prime ($^{\prime}$) denotes differentiation with respect to $r$, to be realized.

\begin{figure}[!htbp]
\begin{centering}
    \includegraphics[width=1\columnwidth]{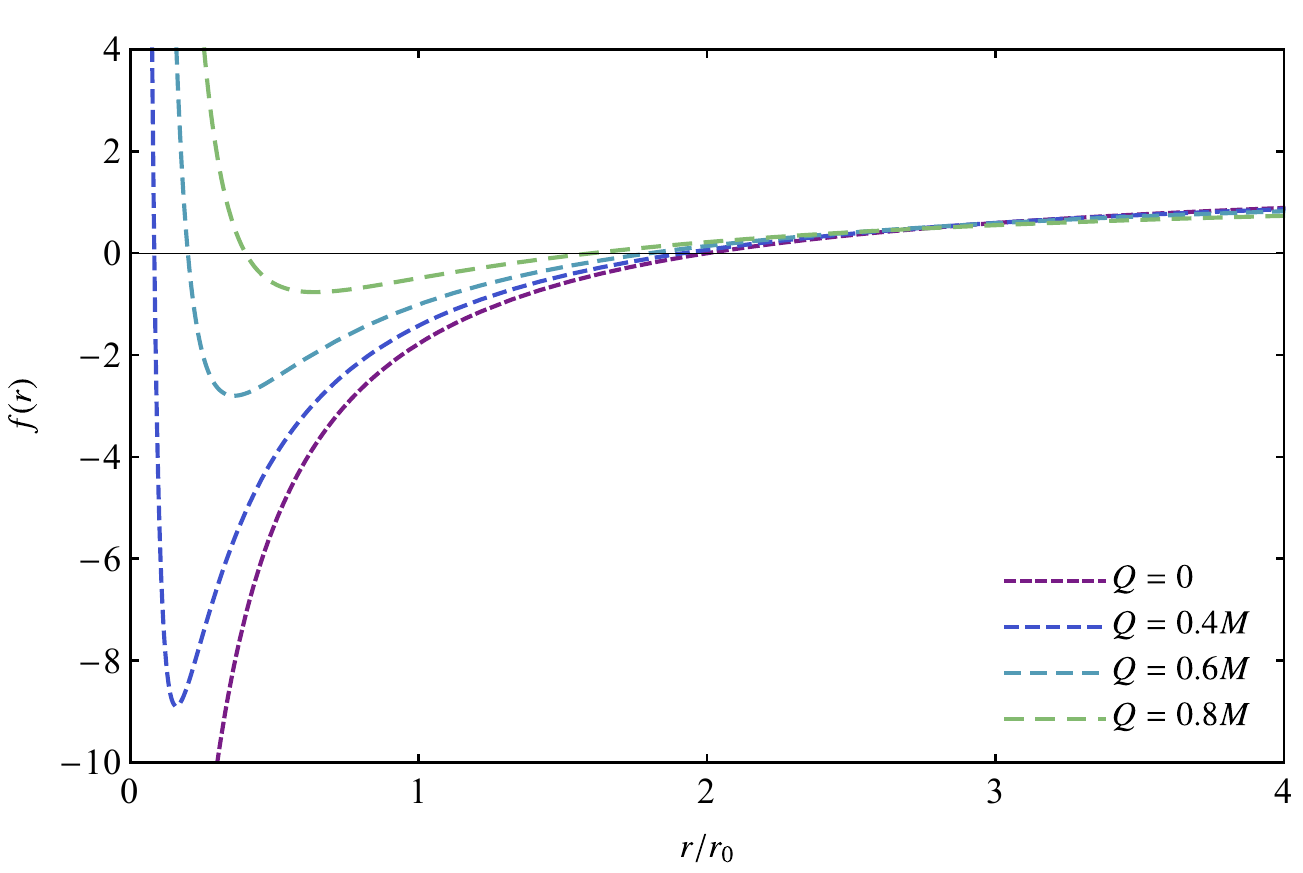}
    \caption{Metric function of the charged BH in LQG, considering distinct values of $Q/M$, as a function of $r/r_{0}$. The corresponding locations of the bounce radius $r_{0}$ are given by $r_{0}/M = 1.77778$, $1.69381$, $1.57454$, and $1.35924$, respectively. For comparison, we exhibit the chargeless counterpart ($Q = 0$), and fixed $b_{0} = 3$.}
    \label{mf}
\end{centering}
\end{figure}

In Fig.~\ref{mf}, we exhibit the metric function of the charged LQG BH geometry normalized by the bounce radius. We observe that BHs exist when the condition $|Q| \leq Q_{\rm{ext}}$ is satisfied. For $|Q| < Q_{\rm{ext}}$ we have up to two horizons, given by the roots of $f(r) = 0$ [cf. Eq.~\eqref{horizonss}]. On the other hand, $|Q| > Q_{\rm{ext}}$ is related to horizonless solutions whose scope is beyond this paper. Moreover, note that for each value of $Q/M$, there is a corresponding bounce radius hidden within the BH but outside the Cauchy horizon, as pointed out in Eq.~\eqref{r0constraint}. We also highlight that due to Eq.~\eqref{ext}, a given BH charge-to-mass ratio such as $Q = 0.9M$ does not lead to a BH solution for $M$ satisfying all values between $0 < M \leq 1$. For this choice of $Q/M$ in particular, Eq.~\eqref{ext} implies that, for there to still be a BH solution, the BH mass must satisfy $M \lesssim 0.873$. 

\begin{figure}[!htbp]
\begin{centering}
    \includegraphics[width=\columnwidth]{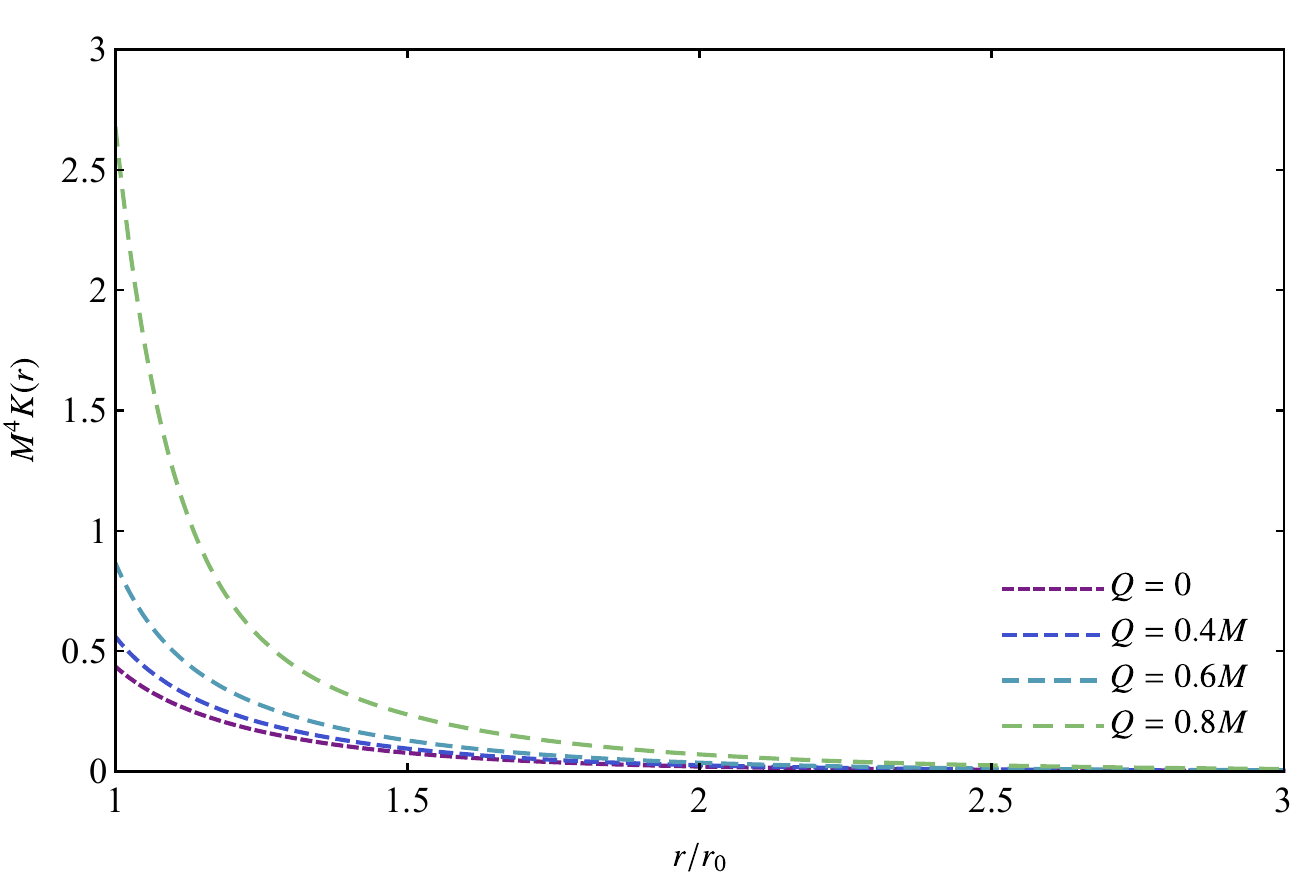}
    \caption{Kretschmann scalar invariant of the charged LQG BH spacetime, as a function of $r/r_{0}$, considering distinct values of $Q/M$. For comparison, we show the chargeless case ($Q = 0$), and set $b_{0} = 3$.}
    \label{kscalar}
\end{centering}
\end{figure}

In Fig.~\ref{kscalar}, we display the Kretschmann scalar, defined as
\begin{equation}
K \equiv R_{\mu\nu\sigma\rho}R^{\mu\nu\sigma\rho},
\end{equation}
where $R_{\mu\nu\sigma\rho}$ is the Riemann tensor for the charged LQG BH spacetime. We observe that the spacetime is regular for $r \geq r_{0}$, in the sense that all invariants constructed from the Riemann tensor and the metric tensor are finite~\cite{Lobo:2020ffi,Bronnikov:2012wsj}.

\section{Geodesic analysis}\label{sec:ga}

In this section, we present the equations of motion for null geodesics in the charged LQG BH spacetime. Due to spherical symmetry, we consider the motion in the equatorial plane without loss of generality. The classical Hamiltonian associated with the motion of massless particles can be written as
\begin{align}
\nonumber \mathrm{H} & \equiv \dfrac{1}{2} g^{\mu\nu}k_{\mu}k_{\nu} \\
\label{L_SG} & = \dfrac{1}{2}\left[-\dfrac{k_{t}^{2}}{f(r)} + f(r)\left( 1 - \dfrac{r_0}{M} g(r) \right)k_{r}^{2} + \dfrac{k_{\varphi}^{2}}{r^{2}}\right],
\end{align}
where $k_\mu$ are the covariant components of the four-momentum of massless particles. Using Hamilton's equations, given by
\begin{equation}
\dot{x}^{\mu} = \dfrac{\partial \mathrm{H}}{\partial k_{\mu}},
\end{equation}
we find that 
\begin{align}
\label{eqm1}\dot{t} &= -\dfrac{k_{t}}{f(r)}  ,\\
\label{eqm2}\dot{r} &= f(r)\left( 1 - \dfrac{r_0}{M} g(r) \right)k_{r}  ,\\
\label{eqm3}\dot{\varphi} &=  \dfrac{k_{\varphi}}{r^{2}}.
\end{align}
Since the Hamiltonian~\eqref{L_SG} does not depend explicitly on the coordinates $t$ and $\varphi$, we have that $k_{t} \equiv -E$ and $k_{\varphi} \equiv L$ are constants of motion, where $E$ and $L$ are the energy and angular momentum of the massless particles, respectively.

\begin{figure}[!htbp]
\begin{centering}
\includegraphics[width=\columnwidth]{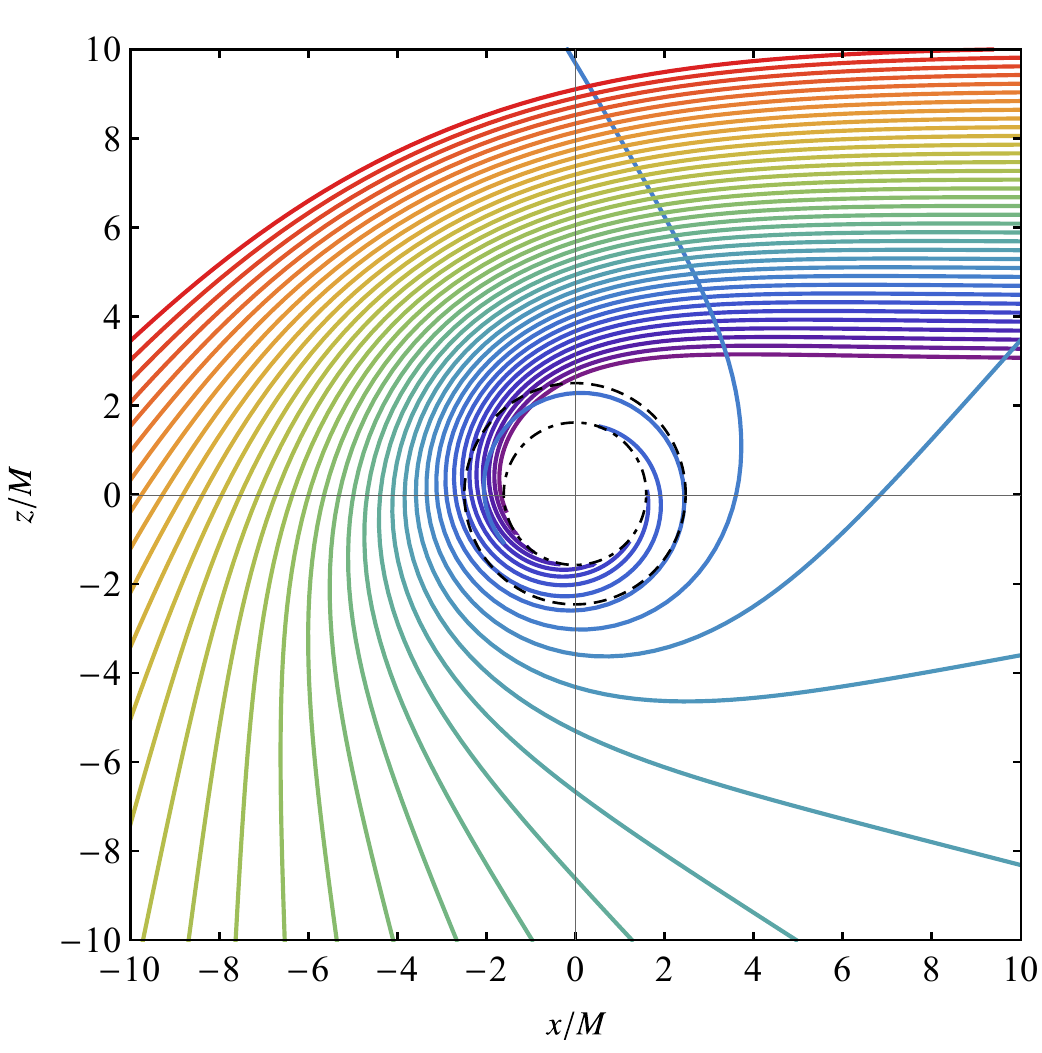}
\caption{Trajectories of massless particles in the charged LQG BH spacetime with $Q = 0.8M$ and $b_{0} = 3$, considering distinct impact parameters. The dashed circle is the unstable circular orbit $r_{c} = 2.4848$, for which the corresponding critical impact parameter is $b_{c} = 4.5459M$. The inner dot-dashed circle is the corresponding event horizon location, i.e., $r_{+} = 1.6M$. The initial conditions are given by $r_{\rm{inf}} = 50M$ and $\varphi = \pi - \arctan(3\sqrt{3}/100)$ at $t_{0} = 0$.}
    \label{geodesics}
\end{centering}
\end{figure}

Using Eqs.~\eqref{eqm1}-\eqref{eqm3} and the condition $\mathrm{H} = 0$, we obtain a radial equation for massless particles, given by
\begin{equation}
\label{ME}\dfrac{\dot{r}^{2}}{L^{2}} \equiv U(r) = \left( 1 - \dfrac{r_0}{M} g(r) \right)\left(\dfrac{1}{b^{2}}-\dfrac{f(r)}{r^{2}} \right),
\end{equation}
where $b \equiv L/E$ is the impact parameter. From 
\begin{equation}
\dot{r}|_{r = r_{c}} = 0 \quad \text{and} \quad \ddot{r}\big|_{r = r_{c}} = 0,
\end{equation}
we find the critical radius $r_{c}$ of the unstable circular orbit and the critical impact parameter $b_{c}$, namely,
\begin{align}
\label{CR}2f(r_{c})-r_{\rm{c}}f'(r_{c}) = 0 \quad \text{and} \quad b_{c} = \dfrac{L_{c}}{E_{c}} = \dfrac{r_{c}}{\sqrt{f(r_{c})}},
\end{align}
respectively. Notice that these equations are the same as those derived for the RN case. In Fig.~\ref{geodesics}, we exhibit some geodesics of massless particles in the background of the charged LQG BH spacetime. We can obtain these trajectories by numerically integrating the radial equation~\eqref{ME} and its first derivative. As we can observe, for $b < b_{c}$, the geodesics are absorbed, while for $b > b_{c}$ they are scattered. On the other hand, the situation $b = b_{c}$ is related to a geodesic moving around the BH in an unstable circular orbit with radius $r_{c}$.

The classical capture cross section of geodesics, also known as the geometric cross section (GCS), is given by~\cite{wald2010general}
\begin{equation}
\label{GCS}\sigma_{\rm{gcs}} \equiv \pi b_{\rm{c}}^{2}.
\end{equation}
Notice that in the spherically symmetric scenario, $b_c$ is the shadow radius $r_{s}$ as seen by a distant observer~\cite{dePaula:2023ozi}. Therefore, the shape of the shadow can be obtained by the parametric plot of Eq.~\eqref{GCS}. Moreover, since the expressions for $f(r)$, $r_c$, and $b_{c}$ of the charged LQG and RN BHs are the same [cf. Eqs.~\eqref{LEfinal} and~\eqref{CR}], these BH solutions have the same GCS and shadow radius for the same choice of $Q/M$. Consequently, any observational test relying purely on the shadow size cannot distinguish the charged LQG BHs from a classical one. This makes investigations considering other physical observables, such as the total ACS and thermodynamic quantities, fundamentally necessary to find quantum signatures.

In the high-energy regime, the ACS can be described by a formula known as the sinc approximation, which takes into account the GCS and some features of null unstable geodesics. This formula helps anticipate some wave interference effects in the vicinity of the BH and can be written as~\cite{D_canini_2011_feb,D_canini_2011_aug}
\begin{equation}
\label{SINC}\sigma_{\rm{hf}} \approx \sigma_{\rm{gcs}}\left[1-8\pi b_{c} \Lambda e^{-\pi b_{c} \Lambda} \text{sinc}(2\pi b_{c} \omega)\right],
\end{equation}
where $\text{sinc}(x) \equiv \sin(x)/x$ and $\Lambda$ is the Lyapunov exponent related to circular null geodesics~\cite{Cardoso:2008bp}, namely,
\begin{equation}
\label{LEUCO}\Lambda = \sqrt{\dfrac{L^{2}}{2\dot{t}^{2}}\left(\dfrac{d^{2}U(r)}{dr^{2}}\right)\bigg|_{r = r_{c}}}.
\end{equation}

\begin{figure}[!htbp]
\begin{centering}
\includegraphics[width=\columnwidth]{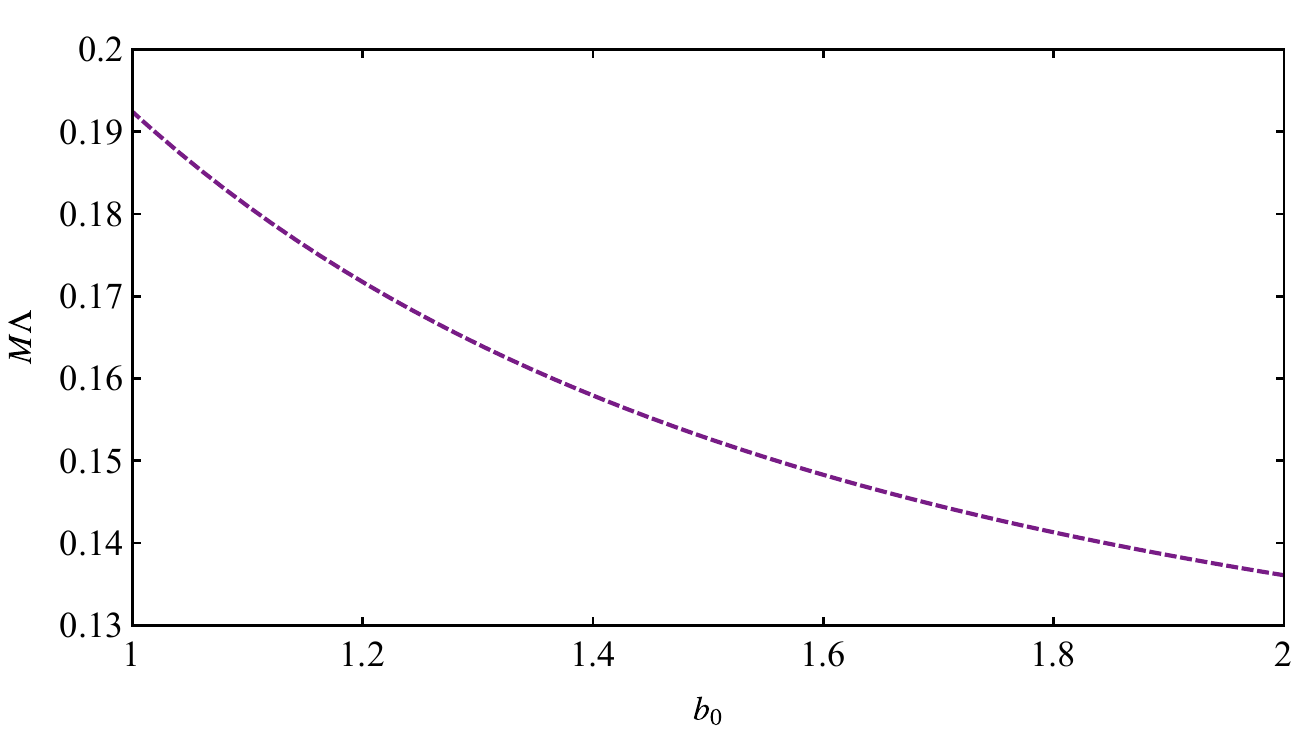}
\caption{Lyapunov exponent of the charged LQG, as a function of $b_{0}$. For simplicity, we set $Q = 0$. The increase of $Q/M$ leads to similar results as those obtained for the RN case; see, for example, Ref.~\cite{Paula:2020yfr}, in particular Fig. 8. Moreover, notice that for $b_{0} = 1$, we obtain the Schwarzschild result given by $M\Lambda = 0.19245$, as expected.}
    \label{lyapunov}
\end{centering}
\end{figure}

In Fig.~\ref{lyapunov}, we display the Lyapunov exponent as a function of the quantum parameter $b_{0}$\footnote{For simplicity, throughout this work, we refer to the quantity $b_{0}$ [cf. Eq.~\eqref{b0def}] as the ``quantum parameter'', since it is this parameter that captures the LQG contributions to the spacetime geometry [cf. Eq.~\eqref{LEfinal}].}. As we can see, the Lyapunov exponent typically decreases as we increase the value of $b_{0}$. Therefore, the instability timescale ($\tau = 1/\Lambda$) gets longer. In Sec.~\ref{sec:na}, we compare the classical [cf. Eq.~\eqref{GCS}] and semiclassical [cf. Eq.~\eqref{SINC}] approximations for the geodesic scattering with the numerical results for the total ACS.

\section{Wave equation}\label{sec:we}

In this section, we investigate the propagation of neutral, massless scalar waves in the vicinity of the charged LQG BH geometry. We also present an expression for the total ACS obtained via the partial-wave method. For completeness, we also discuss analytical bounds for the GBF.

\subsection{Scalar wave}\label{subsec:sw}

The Klein-Gordon equation that governs the dynamics of a neutral, massless test scalar field $\Phi$ in the background of curved spacetimes is given by
\begin{equation}
\label{KG}\dfrac{1}{\sqrt{-g}}\partial_{\mu}\left(\sqrt{-g}g^{\mu \nu}\partial_{\nu}\Phi\right) = 0.
\end{equation}
In view of spherical symmetry, and assuming the incident plane wave propagates along the polar axis, we can decompose $\Phi$ as
\begin{equation}
\label{PHI}\Phi= \dfrac{1}{r}\sum_{l}^{\infty} C_{\omega l}\Psi_{\omega l}(r)P_{l}(\cos\theta)e^{-i\omega t},
\end{equation}
where $C_{\omega l}$ are coefficients, with $\omega$ and $l$ being the frequency and angular momentum of the scalar field, respectively. The function $P_{l}$ is the Legendre polynomial, and $\Psi_{\omega l}$ are the radial functions. Using the tortoise coordinate $r_{\star}$, defined as
\begin{equation}
dr_{\star}\equiv\dfrac{dr}{f(r)\sqrt{1 - r_{0}g(r)/M}},
\end{equation}
we can show that $\Psi_{\omega l}$ satisfies 
\begin{equation}
\label{RE_TC}\frac{d^{2}}{dr_{\star}^{2}}\Psi_{\omega l}+\left[\omega^{2}-V(r)\right]\Psi_{\omega l}=0,
\end{equation}
where the effective potential $V(r)$ is given by
\begin{equation}
\label{EP}V(r) = \dfrac{h(r)f^{\prime}(r)}{2r} + f(r)\left(\dfrac{h^{\prime}(r)}{2r}+\dfrac{l(l+1)}{r^{2}}\right),
\end{equation}
with $h(r) = f(r)(1 - r_{0}g(r)/M)$ for the charged LQG BH.

\begin{figure}[!htbp]
\begin{centering}
    \includegraphics[width=1\columnwidth]{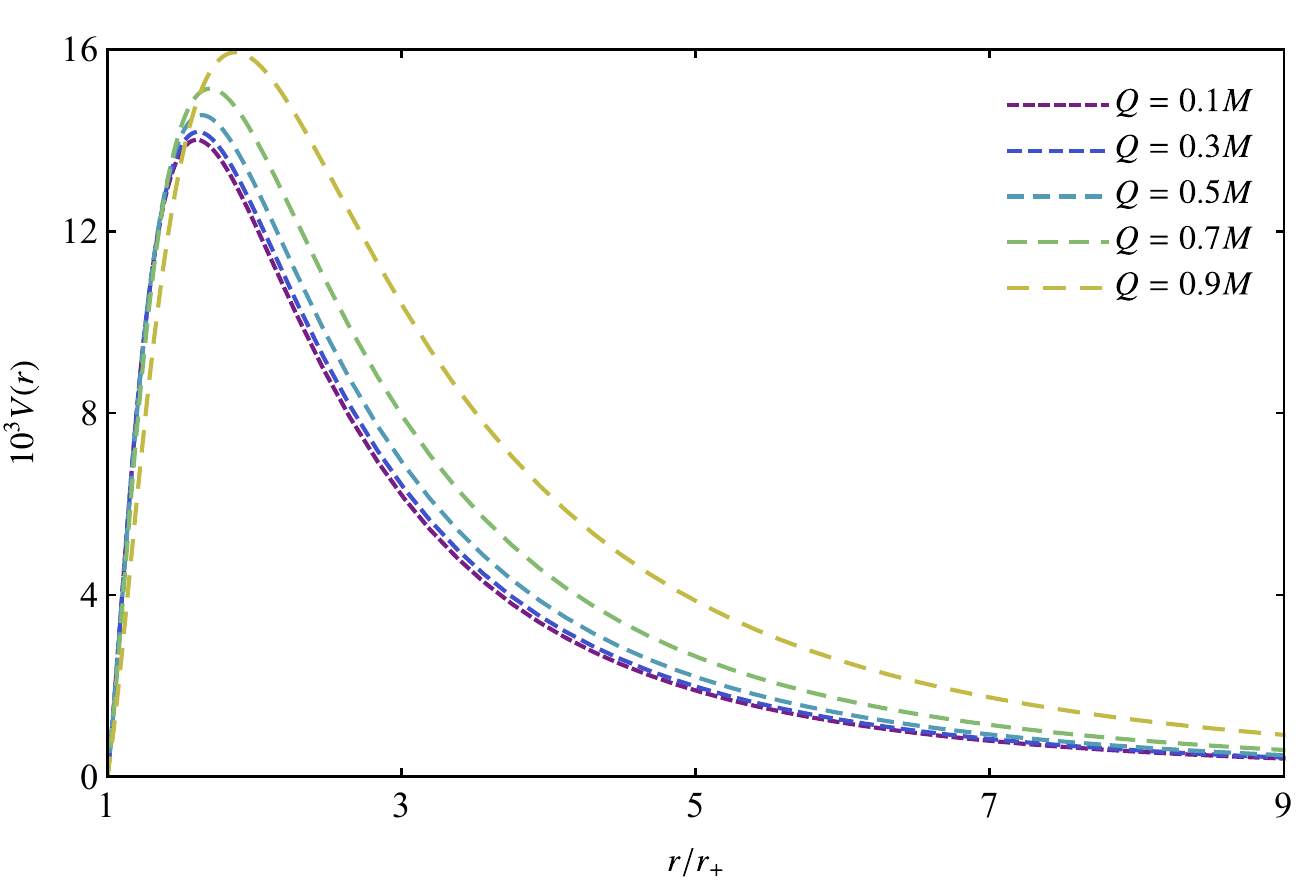}    \includegraphics[width=1\columnwidth]{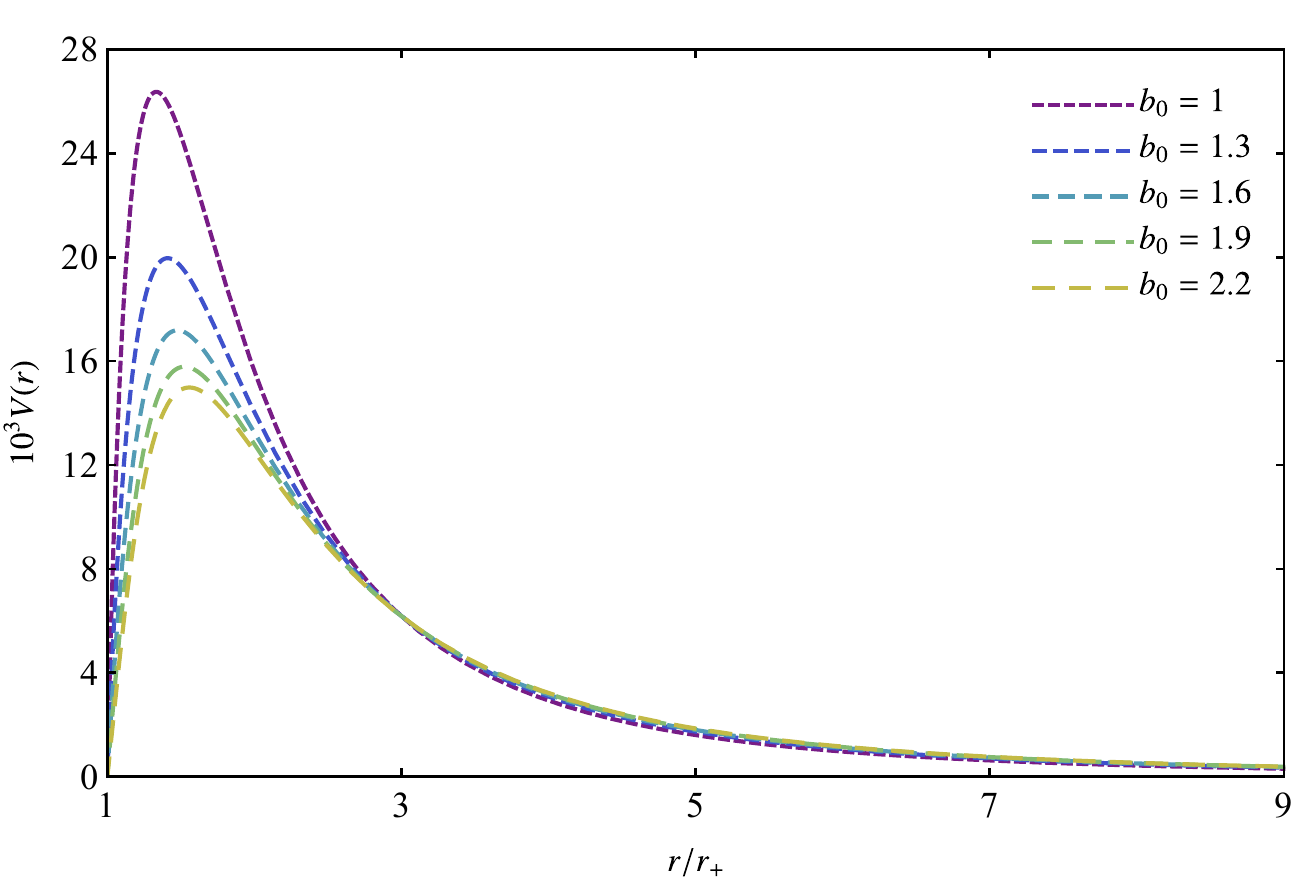}
    \caption{Effective potential of massless scalar waves in the background of the charged LQG BH spacetime, as a function of $r/r_{+}$, considering two distinct scenarios: (i) for distinct values of $Q/M$ with $b_0 = 3$ (top panel); and (ii) for different choices of $b_{0}$ with $Q = 0.5M$ (bottom panel). Here, we set $l = 0$ for all cases.}
    \label{VdifQ}
\end{centering}
\end{figure}

In Fig.~\ref{VdifQ}, we display the effective potential of massless scalar waves in the background of the charged LQG BH spacetime. As we can see, the maximum of the effective potential increases (decreases) as we consider higher values of $Q/M$ ($b_0$) for fixed values of $b_{0}$ ($Q/M$). In situations where the potential barrier decreases (increases), the scalar waves are expected to be more (less) absorbed. Notice that the centrifugal term of the effective potential, namely, $l(l+1)/r^{2}$, dominates the behavior of the potential barrier for higher values of $l$, increasing the peak of the effective potential. Moreover, it is important to emphasize that because $r_0 < r_+$, the incident massless scalar waves never reach the bounce radius. 

The solutions of the Klein-Gordon equation consistent with the absorption/scattering problem are given by
\begin{equation}
\label{BC}\Psi_{\omega l}\sim\begin{cases}
T_{\omega l}e^{-i\omega r_{\star}}, & r_{\star}\rightarrow -\infty  \ (r\rightarrow r_{+}),\\
e^{-i\omega r_{\star}}+R_{\omega l}e^{i\omega r_{\star}}, & r_{\star}\rightarrow \infty \ (r\rightarrow \infty),
\end{cases}
\end{equation}
where $T_{\omega l}$ and $R_{\omega l}$ are complex coefficients that satisfy
\begin{equation}
\label{CF}|R_{\omega l}|^{2}+|T_{\omega l}|^{2} = 1,
\end{equation}
due to the conservation of the flux. The solutions given by Eq.~\eqref{BC} can be interpreted as follows. We have an ingoing wave heading towards the BH from spatial infinity ($e^{-i\omega r_{\star}}$) that, when interacting with the effective potential given by Eq.~\eqref{EP}, is partially scattered back to spatial infinity ($e^{i\omega r_{\star}}$) and partially absorbed by (transmitted into) the BH ($e^{-i\omega r_{\star}}$).

\subsection{ACS and GBF}\label{subsec:acsgbf}

It is usual to obtain an expression for the total ACS $\sigma$ as a sum of contributions from partial waves $\sigma_{l}$. For that purpose, we can follow a standard procedure. We expand the scalar field as the sum of asymptotic plane waves. Then we fix $C_{\omega l}$ with appropriate boundary conditions; see Eqs.~\eqref{BC}. Thus, we get~\cite{Unruh:1976fm,Paula:2020yfr}
\begin{equation}
\label{ACS}\sigma = \sum_{l = 0}^{\infty}\sigma_{l},
\end{equation}
where $\sigma_{l}$ is given by
\begin{equation}
\label{PACS}\sigma_{l} = \dfrac{\pi}{\omega^{2}}(2l+1)|T_{\omega l}|^{2}.
\end{equation}

The GBF corresponds to the probability of an incoming wave being absorbed by the BH, i.e., the absorption probability. Indeed, these factors can be defined as~\cite{Pedrotti:2025idg}
\begin{equation}
\label{gbf}\Gamma_l (\omega) = |T_{\omega l}|^{2}.
\end{equation}
Therefore, the GBF and the ACS are closely connected. In addition to that, it is possible to obtain semi-analytical upper bounds for the GBF~\cite{Boonserm:2008zg,Sakalli:2022xrb,Nakarachinda:2025hcd}. According to Refs.~\cite{Boonserm:2008qf,Boonserm:2009zba}, the general (analytical) bound for the GBF can be written as
\begin{equation}
\label{gbfapprox}\Gamma^{\rm{ana}}_{l}(\omega) \geq \sech^{2}\left[\int_{-\infty}^{\infty}\vartheta dr_{\star} \right],
\end{equation}
in which 
\begin{equation}
\vartheta = \dfrac{1}{2z(r_{*})}\sqrt{z^{\prime}(r_{*})^{2}+\left[\omega^{2}  - V(r) - z(r_{*})^{2} \right]^2}.
\end{equation}
Moreover, $z(r_{\star})$ is a positive function, i.e., $z(r_{\star}) > 0$, satisfying $z(-\infty) = z(\infty) = \omega$.

\begin{figure}[!htbp]
\begin{centering}
    \includegraphics[width=1\columnwidth]{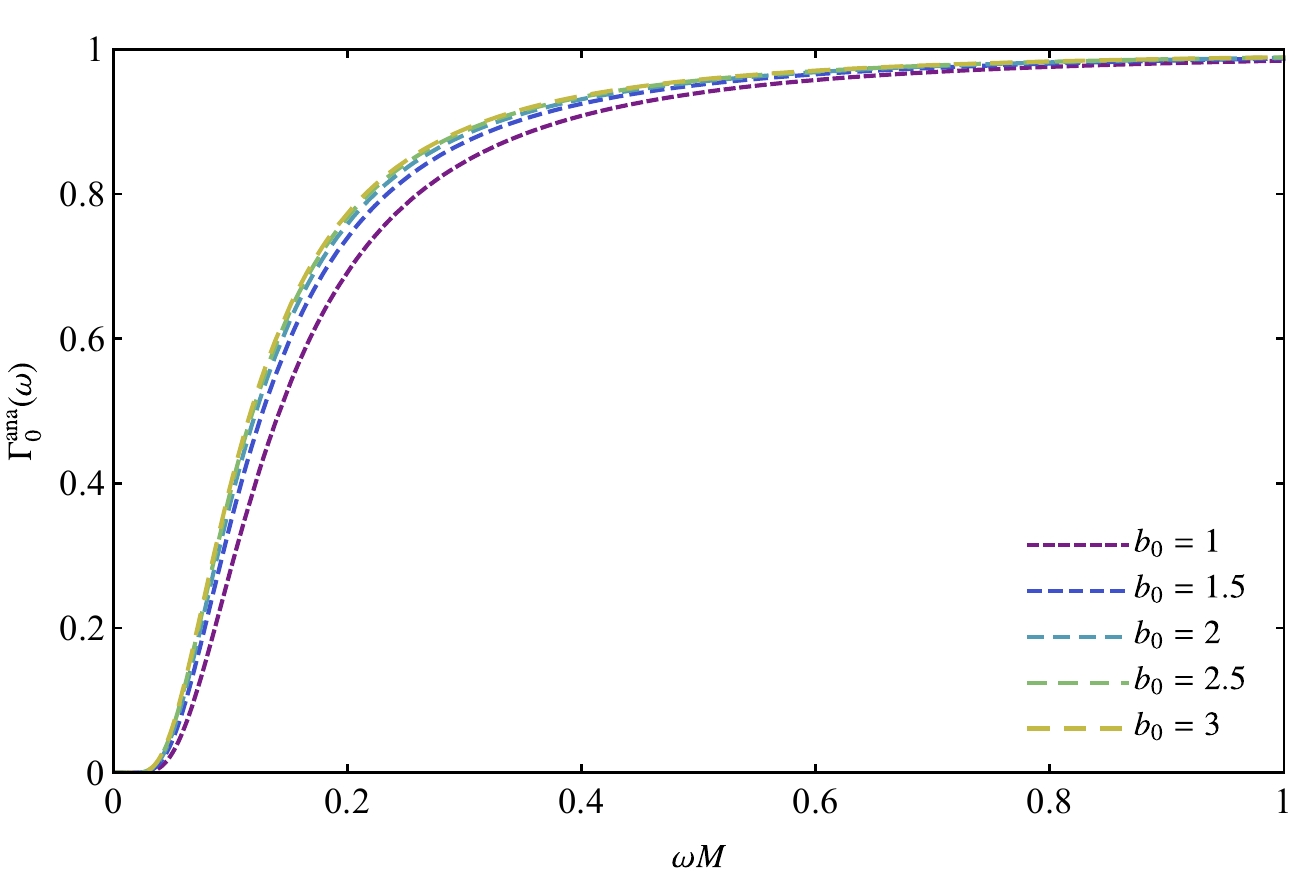}
    \caption{GBF computed analytically, as a function of $\omega M$, considering distinct values of $b_{0}$, for $Q = 0$ and $l = 0$.}
    \label{gbfana}
\end{centering}
\end{figure}

Notice that solving Eq.~\eqref{gbfapprox} for a complex function $z(r_{*})$, considering the effective potential given by Eq.~\eqref{EP}, can be quite complicated. To simplify the analytical treatment, we consider the special case of $z = \omega$ and $Q = 0$. This is enough to obtain useful practical results~\cite{Boonserm:2008zg}. Thus, we find that
\begin{equation}
\label{gbfanal}\Gamma^{\rm{ana}}_{l}(\omega) \geq \sech^{2} \left(X\right),
\end{equation}
where
\begin{equation}
X = \dfrac{b_{0} (3 b_{0} (5 b_{0} l (l+1)+b_{0}+5 l (l+1)+2)+4)+2}{30 b_{0} (b_{0}+1)^2 M \omega }.
\end{equation}
In Fig.~\ref{gbfana}, we display this approximation for different values of $b_{0}$. As we can see, in the low-frequency regime, the wave is completely reflected by the potential barrier, i.e., $\Gamma^{\rm{ana}}_{l}(\omega) \rightarrow 0$. Nevertheless, as the frequency increases, the wave tunnels the potential barrier and starts to be absorbed. Finally, when the frequency reaches a threshold, the wave is completely transmitted. We also observe that GBFs tend to increase as we consider higher values of $b_{0}$, which means that as $b_{0}$ increases, more of the incoming wave gets absorbed. This is consistent with the behavior of the effective potential (see the bottom panel of Fig.~\ref{VdifQ}). In Sec.~\ref{sec:na}, we compare the analytical approximation given by Eq.~\eqref{gbfanal} with the numerics.

\section{Main results}\label{sec:mr}

In this section, we present a selection of our main results concerning the absorption spectrum of massless scalar waves impinging on the charged LQG BH. For simplicity, we divide this section into four parts: (i) numerical approach, (ii) GBF, (iii) absorption spectrum, and (iv) mimic setups.

\subsection{Numerical approach}\label{sec:na}

We apply the stiffness switching integration method~\cite{press2007numerical} to numerically solve Eq.~\eqref{RE_TC} from near the event horizon, i.e., $r_{\rm{ini}} = 1.0001 r_{+}$, to far away from the BH, i.e., $r_{\rm{inf}} = 10^{3}M$. The goal is to find numerical values for $\Gamma_{l}(\omega)$. This can be achieved by matching the numerical solutions of Eq.~\eqref{RE_TC} with the appropriate solutions to the absorption problem, given by Eq.~\eqref{BC}. The linear system relating the radial function and its first derivative at numerical infinity with the complex coefficients can be written as
\begin{align}
\label{LS}\begin{bmatrix}
\Psi_{\omega l}(r) \\ 
\Psi_{\omega l}^{\prime}(r)
\end{bmatrix} 
=
\begin{bmatrix}
e^{-i\omega r_{\star}} & e^{i\omega r_{\star}} \\ 
\left(e^{-i\omega r_{\star}}\right)^{\prime} & \left(e^{i\omega r_{\star}}\right)^{\prime}
\end{bmatrix}\cdot
\begin{bmatrix}
1/T_{\omega l} \\ 
R_{\omega l}/T_{\omega l}
 \end{bmatrix}.
\end{align}
For simplicity, we also normalize the coefficients of Eq.~\eqref{BC} by $T_{\omega l}$. Notice that to properly solve the ordinary differential equations numerically, we need to impose two initial conditions. For our purposes, we fix the field and its first derivative with respect to $r$ near the event horizon as
\begin{subequations}
\begin{align}
\label{bc1}\Psi_{\omega l}(r_{\rm{ini}}) & =  1, \\ 
\label{bc2}\Psi_{\omega l}^{\prime}(r)\big|_{r_{\rm{ini}}} & = -\dfrac{i\omega}{f(r_{\rm{ini}})\sqrt{\left( 1 - \dfrac{r_0}{M} g(r_{\rm{ini}}) \right)}}.
\end{align}
\end{subequations}
The numerical values for $R_{\omega l}$ and $T_{\omega l}$ are found by solving the linear system given by Eq.~\eqref{LS}. With these, we can finally compute the GBF and the total ACS.

\begin{figure}[!htbp]
\begin{centering}
    \includegraphics[width=1\columnwidth]{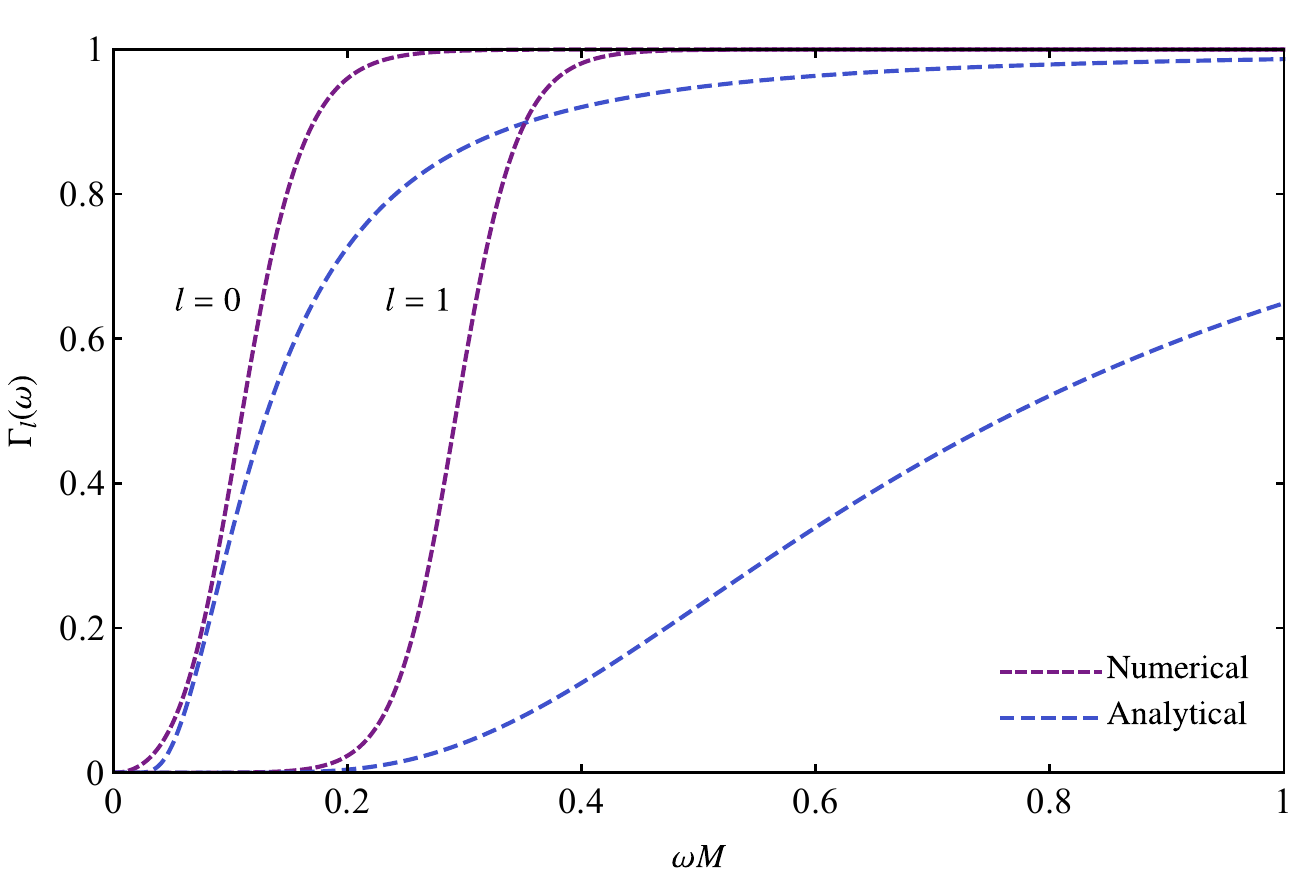}
    \caption{Comparison between the GBFs computed analytically and numerically, as functions of $\omega M$, considering $Q = 0$ and $b_{0} = 3$.}
    \label{gbfcomp}
\end{centering}
\end{figure}

In Fig.~\ref{gbfcomp}, we compare the GBF obtained numerically with the analytical expression given by Eq.~\eqref{gbfanal}. As we can see, the approximation works well for very small frequency values. However, outside this frequency range, the approximation leads to results that are quite different from the numerical ones. Hence, the bound method used in Sec.~\ref{subsec:acsgbf} is useful for dealing with the problem analytically but fails to provide the exact expected values, as also discussed in Ref.~\cite{Nakarachinda:2025hcd}.

\begin{figure}[!htbp]
\begin{centering}
    \includegraphics[width=1\columnwidth]{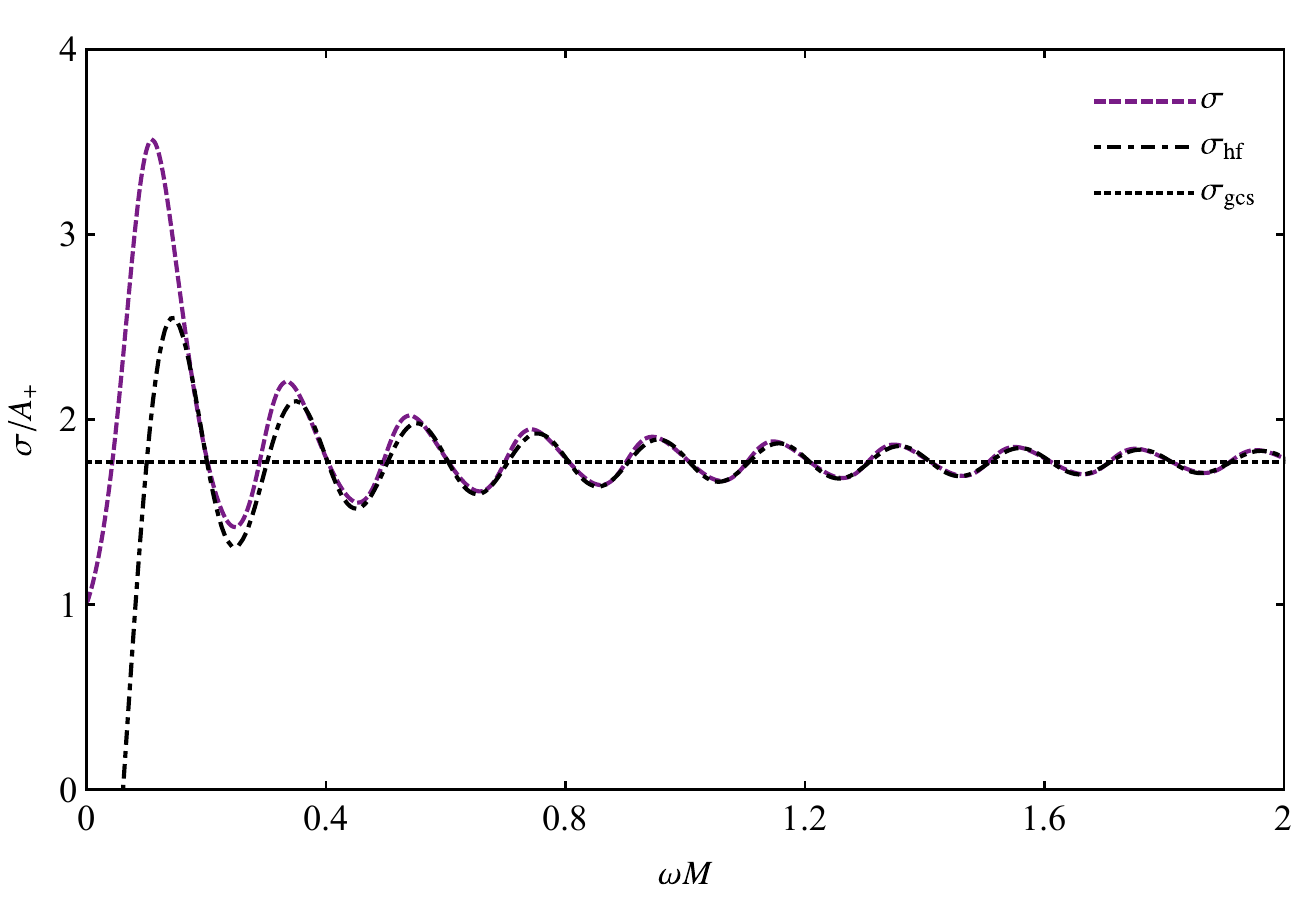}
    \caption{Comparison between the total ACS computed numerically ($\sigma$), the sinc approximation ($\sigma_{\rm{hf}}$), and the GCS ($\sigma_{\rm{gcs}}$), as functions of $\omega M$, considering $Q = 0.5M$ and $b_{0} = 2$.}
    \label{na}
\end{centering}
\end{figure}

In Fig.~\ref{na}, we compare the total ACS computed numerically with the classical and semiclassical approximations discussed in Sec.~\ref{sec:ga}. For simplicity, we display the total ACS normalized by the BH area, which is given by
\begin{equation}
\label{area}A_+ = 4\pi r^{2}_{+}.
\end{equation}
As we can see, the ACS matches the BH area for low-frequency massless scalar waves, as expected~\cite{higuchi2001low}. We also note that the total ACS oscillates around the GCS and the sinc approximation captures the oscillatory pattern of the total ACS for intermediate-to-high frequency ranges.

The results presented in this section show that our numerical results are consistent with the well-known analytical approximations in their corresponding frequency regimes.

\subsection{Greybody factor}\label{subsec:gbf}

\begin{figure}[!htbp]
\begin{centering}
    \includegraphics[width=1\columnwidth]{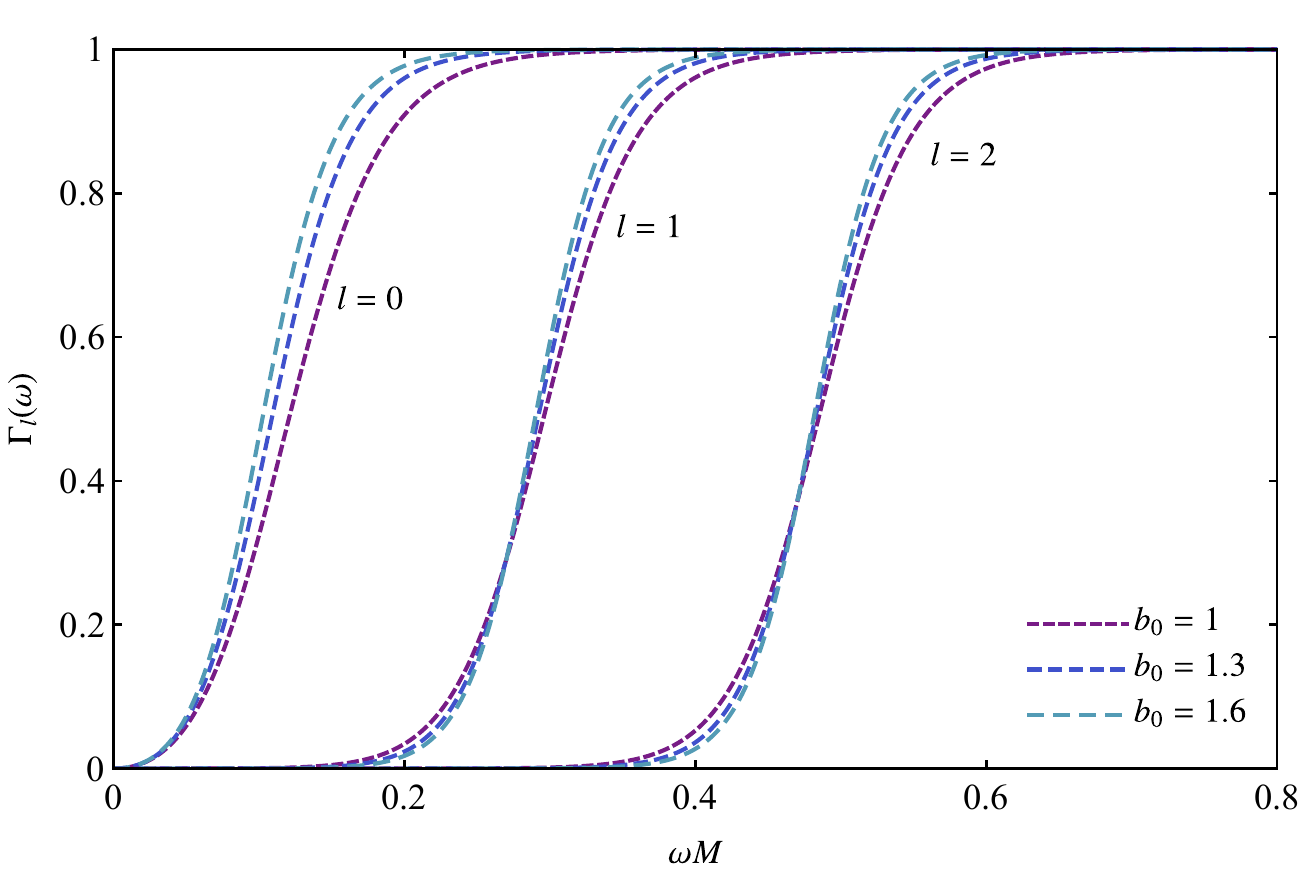}
    \caption{GBFs of massless scalar waves in the background of the charged LQG BH spacetime, as functions of $\omega M$, for distinct values of $l$ and $b_{0}$ with $Q = 0$.}
    \label{gbf0}
\end{centering}
\end{figure}

\begin{figure}[!htbp]
\begin{centering}
    \includegraphics[width=1\columnwidth]{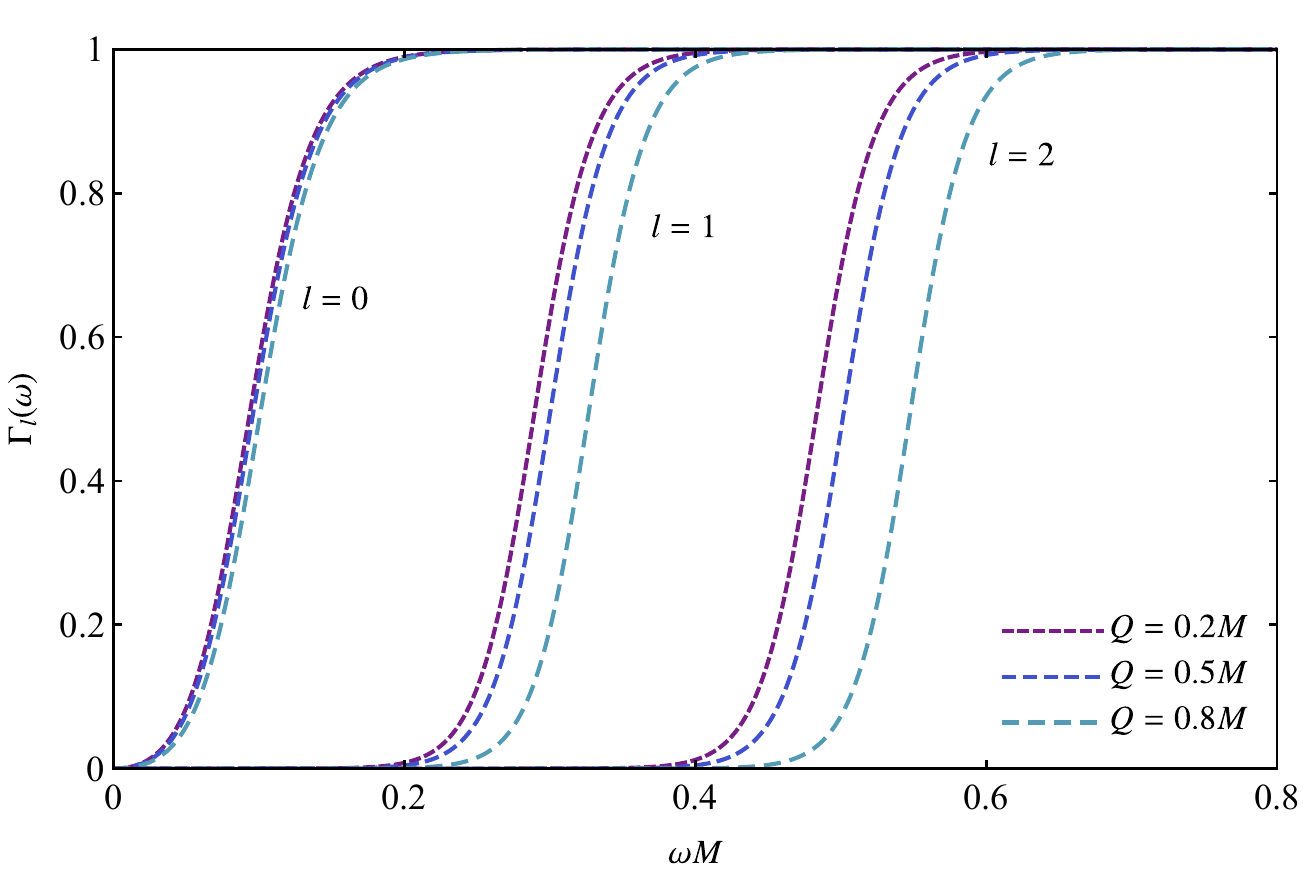}    
    \caption{GBFs of massless scalar waves in the background of the charged LQG BH spacetime, as functions of $\omega M$, for different choices of $l$ and $Q/M$ with $b_{0} = 3$. }
    \label{gbf2}
\end{centering}
\end{figure}

In Figs.~\ref{gbf0} and~\ref{gbf2}, we show the GBFs of massless scalar waves in the background of the charged LQG BH spacetime. We observe that the increase of $b_{0}$ typically increases, for fixed $\omega M$, the GBFs for the fundamental mode $l = 0$, as anticipated by the analytical approximation~\eqref{gbfanal}. However, for $l \geq 1$, the behavior of the GBF is more subtle. As we can see, in a given frequency range, the opposite behavior is observed. Therefore, for $l \geq 1$, considering higher values of $b_{0}$ not only increases the GBFs, as occurs for the mode $l = 0$, but it can also decrease the GBFs, depending on the frequency range. This result can be interpreted as an imprint of (charged) LQG BHs in the GBFs associated with scalar fields.

In Fig.~\ref{gbf2}, we exhibit the GBF for different values of $Q/M$, fixing the value of $b_{0}$. For a fixed value of $l$, we observe a clear ordering as we change the value of the BH charge-to-mass ratio. Moreover, we note that the GBF increases as $Q/M$ increases for a fixed value of the scalar wave frequency. This is what is observed in several BH scenarios~\cite{Paula:2020yfr, Crispino:2009ki, dePaula:2023muc}.

\subsection{Absorption spectrum}\label{subsec:as}

\begin{figure}[!htbp]
\begin{centering}
    \includegraphics[width=1\columnwidth]{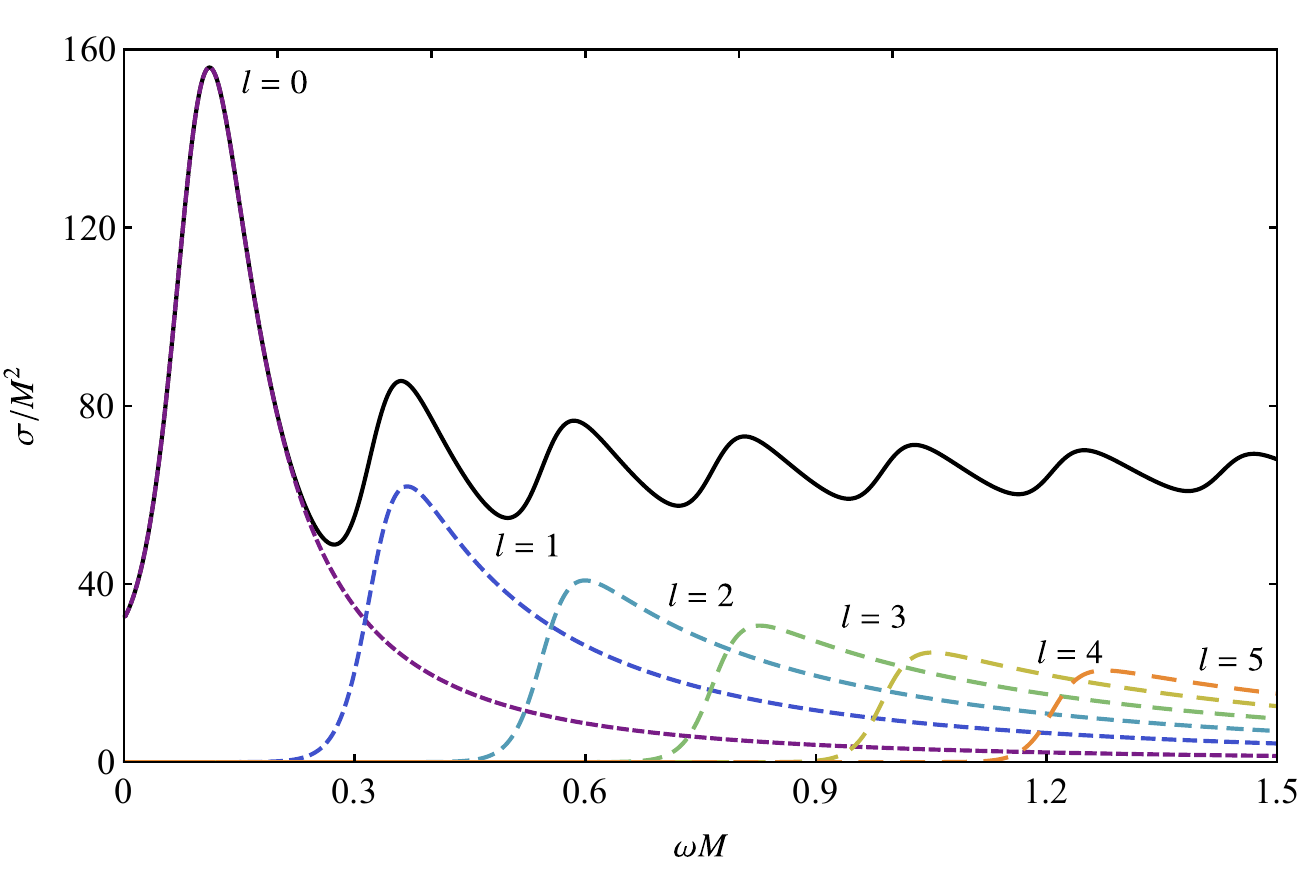}
    \caption{Comparison between the total and partial ACSs of massless scalar waves in the background of the charged LQG BH spacetime, as functions of $\omega M$, considering $Q = 0.8M$ and $b_0 = 3$. }
    \label{compatandpartial}
\end{centering}
\end{figure}

In Fig.~\ref{compatandpartial}, we present the total and partial ACSs for the charged LQG BH. We see that the total ACS is the sum of the partial-wave contributions, as indicated in Eqs.~\eqref{ACS} and~\eqref{PACS}. Moreover, the monopole mode $l = 0$ provides the main contributions to the cross section in the low-frequency regime ($\omega M \rightarrow 0$). Notice also that the oscillatory pattern exhibited by the total ACS can be attributed to the interference of massless scalar waves in the vicinity of the BH.

\begin{figure}[!htbp]
\begin{centering}
    \includegraphics[width=1\columnwidth]{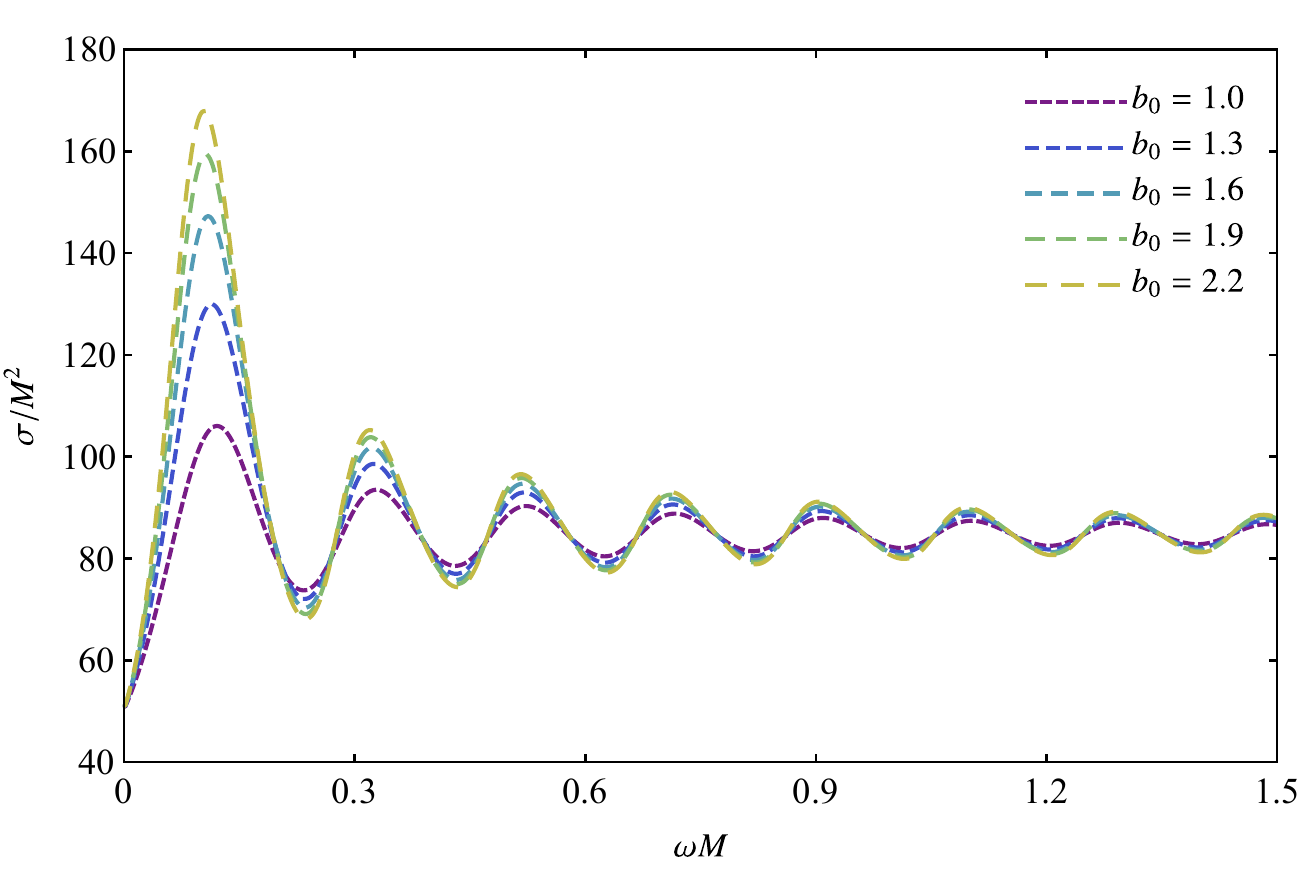}        \includegraphics[width=1\columnwidth]{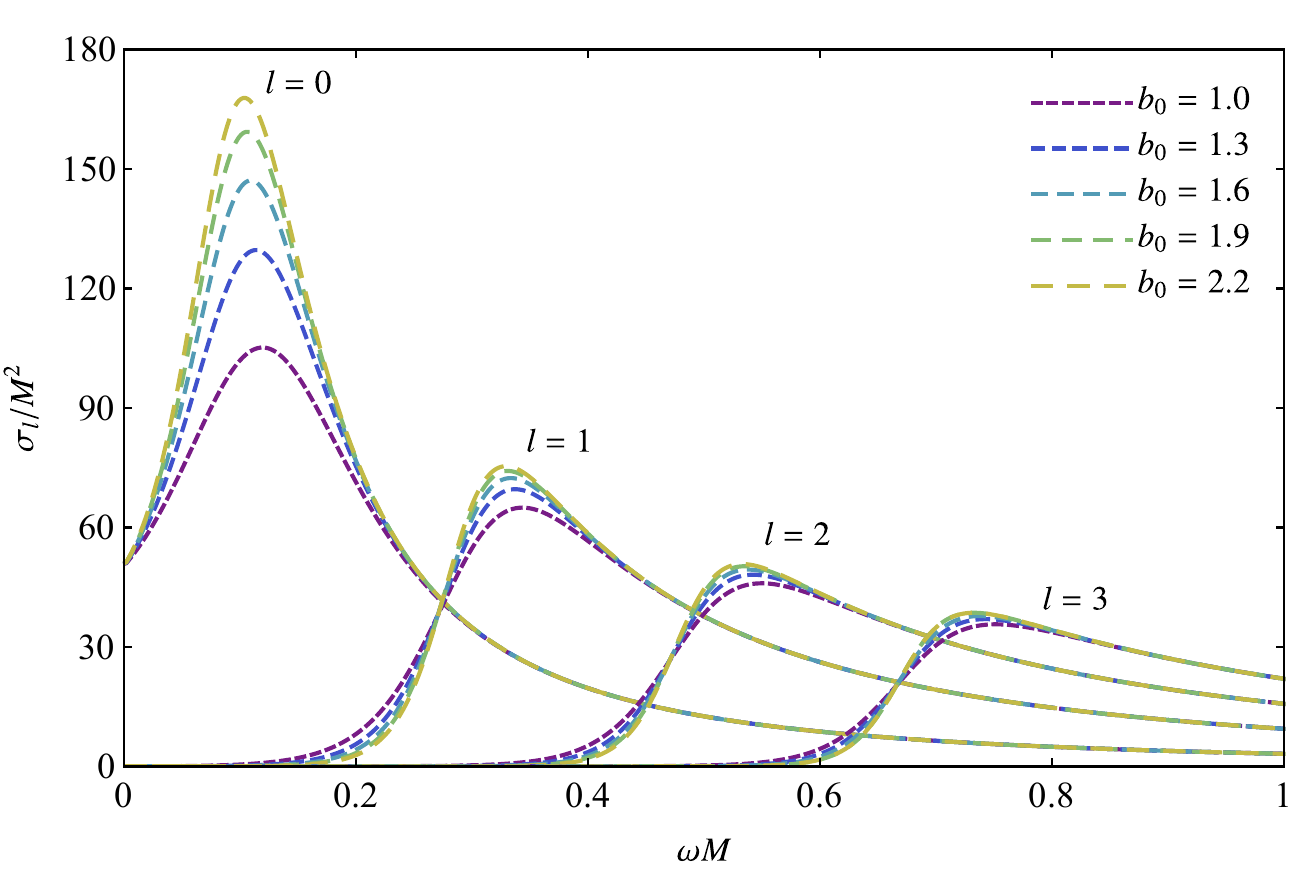}
    \caption{Total (top panel) and partial (bottom panel) ACSs of massless scalar waves in the background of the charged LQG BH, as functions of $\omega M$, for distinct values of $b_{0}$. For simplicity, we set $Q = 0$. }
    \label{tacsdifb}
\end{centering}
\end{figure}

In Fig.~\ref{tacsdifb}, we compare the partial and total ACSs of the charged LQG BH spacetime considering different values of $b_{0}$. We note that the total ACS increases from near the low-frequency regime to the moderate-frequency regime as we consider higher values of $b_{0}$ for a fixed ratio $Q/M$, but tends to the same values in the low- and high-frequency regimes. This can be understood by noting that the quantum parameter $b_{0}$ does not affect the horizon area [cf. Eqs.~\eqref{horizonss} and~\eqref{area}] and the shadow structure [cf. Eqs.~\eqref{CR}] of the charged LQG BH. Hence, we expect the total ACSs to tend to the same values in the low- and high-frequency regimes, since in these regimes we observe that the total ACSs lead to the BH area and the GCS, respectively. Moreover, as discussed in Fig.~\ref{VdifQ}, the peak of the effective potential decreases as we increase the values of $b_0$. Since the lower the potential barrier, the easier it is for the wave to be transmitted to the BH, this explains why the total ACS usually increases as we increase the values of $b_0$. 

\begin{figure}[!htbp]
\begin{centering}
    \includegraphics[width=1\columnwidth]{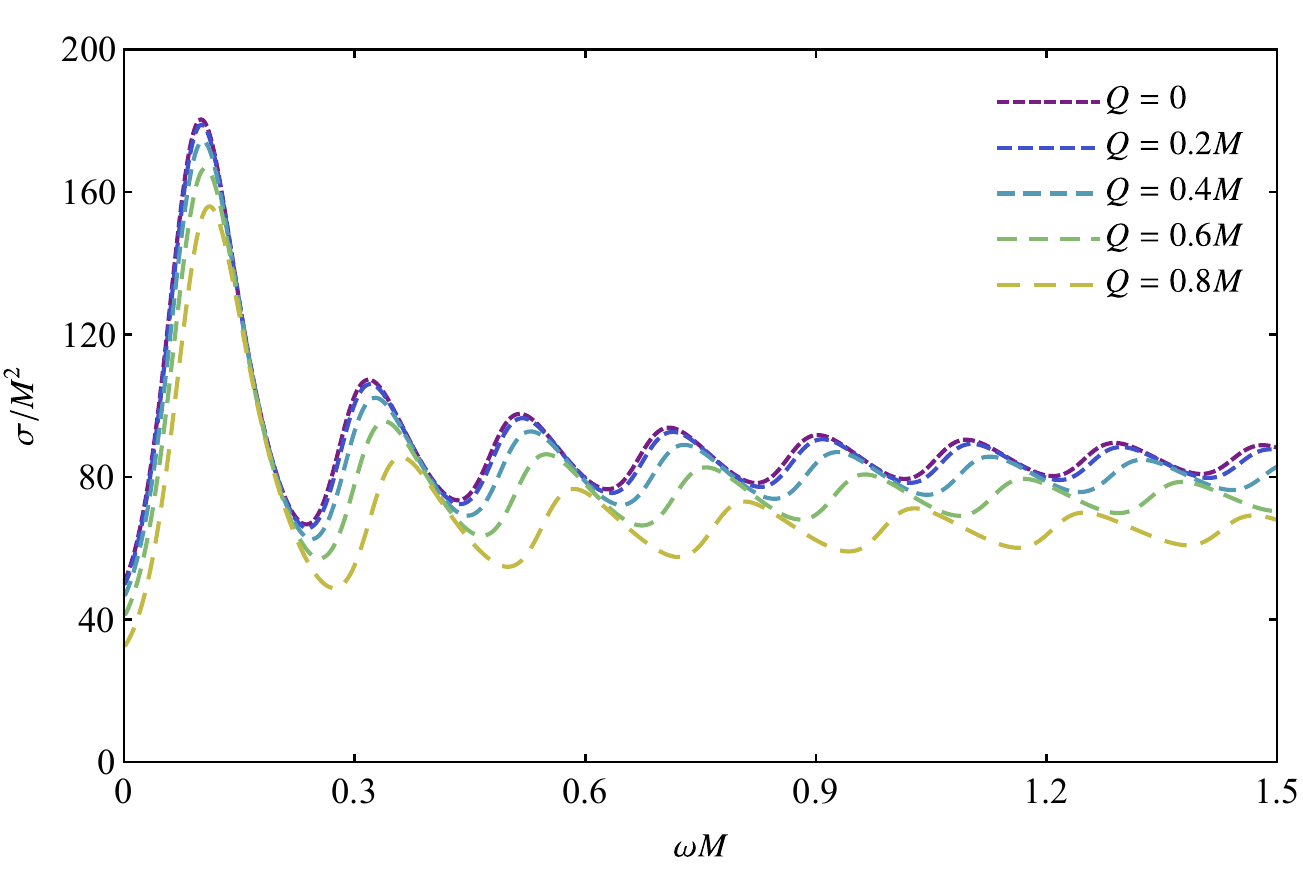}
    \caption{Total ACS of massless scalar waves in the background of the charged LQG BH, as a function of $\omega M$, for different values of $Q/M$. Here, we set $b_{0} = 3$. }
    \label{tacsdifQ}
\end{centering}
\end{figure}

In Fig.~\ref{tacsdifQ}, we display the total ACS of the charged LQG BH spacetime, considering different values of $Q/M$. We observe that the total ACS decreases as we consider higher values of $Q/M$ for a fixed $b_{0}$. This behavior of the total ACS is also consistent with the effective potential, as the potential barrier increases as we increase the values of $Q/M$. Note also that, in this case, we could expect the low- and high-frequency regimes of the total ACS to be different, since we are varying $Q/M$, which affects the horizon area and shadow of the BH.

We finish this section by noting that the behavior of the absorption spectrum as we increase the parameter $b_{0}$ indicates that the peaks and troughs of the total ACS exhibit opposite behaviors. The curve associated with the highest peak corresponds to the deepest troughs. This is in stark contrast with the behavior of the absorption spectra as we consider higher values of $Q/M$ and can play an interesting role in the scattering cross section, which we aim to investigate elsewhere. In addition to that, this unveils a distinct signature of (charged) LQG BHs in the absorption spectrum of scalar fields.    

\subsection{Mimic setups}\label{subsec:ms}

One can seek configurations in which other BH solutions, such as regular or LQG-inspired BHs, and standard BHs, such as RN, have similar physical properties. Concerning the absorption spectrum, a good strategy to find situations in which the ACSs of LQG and RN BHs coincide is to search for the so-called fine-tuned degenerate shadows~\cite{dePaula:2023ozi}. In our case, for asymptotic observers, this corresponds to
\begin{equation}
\label{similar}b_{c}^{\rm{RN}}(Q_{\rm{RN}}) = b_{c}^{\rm{LQG}}(Q_{\rm{LQG}},b_{0}).
\end{equation}
Notice that for the charged LQG BH spacetime considered in this work, the relation~\eqref{similar} is always satisfied since the shadow radius of the LQG and RN BHs coincide. Therefore, the constraint on $Q/M$ is given by Eq.~\eqref{constraintQ}. For example, considering $b_0 = 1.7$, we have LQG and RN BHs that satisfy Eq.~\eqref{similar} for $Q_{\rm{LQG}} \lesssim 0.80869M$; otherwise, Eq.~\eqref{constraintQ} is violated.

\begin{figure*}[!htbp]
\begin{centering}
    \includegraphics[width=1\columnwidth]{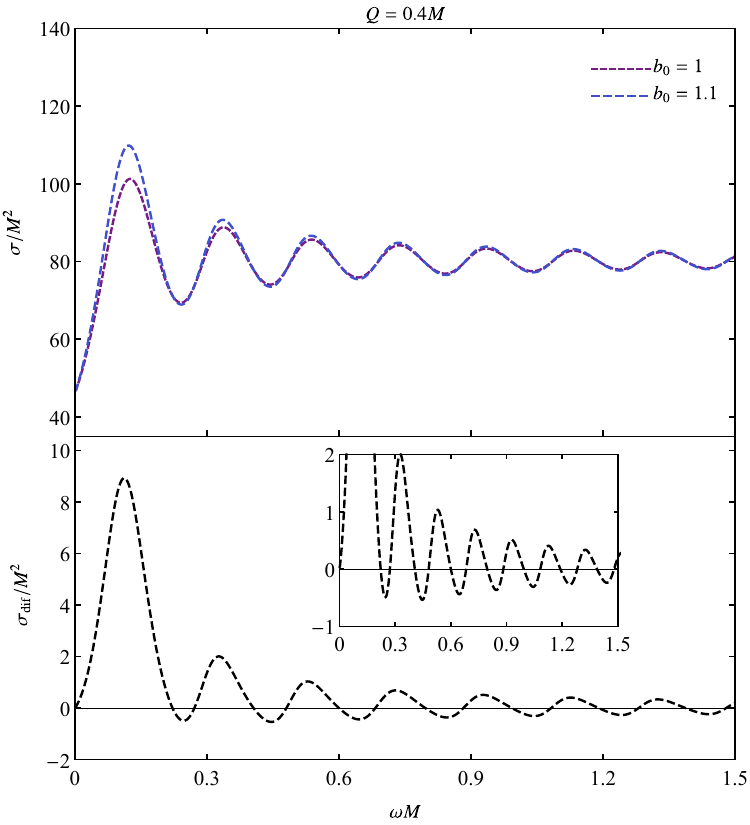} \quad
    \includegraphics[width=1\columnwidth]{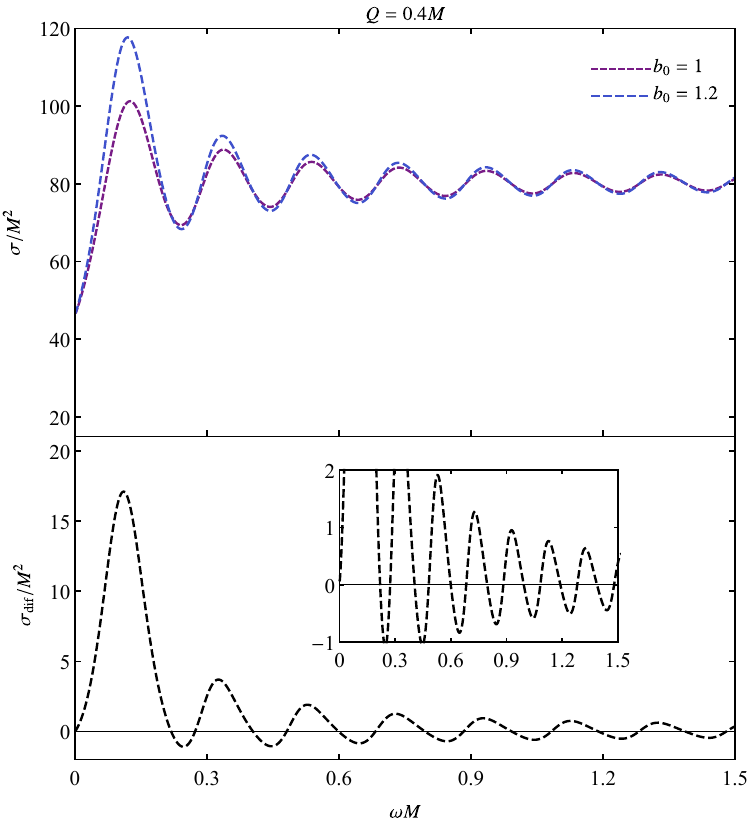}
    \caption{Total ACSs of massless scalar waves in the background of the charged LQG and RN BH spacetimes with $Q = 0.4M$, as functions of $\omega M$, considering two distinct configurations: (i) $b_{0} = 1.1$ (top-left panel); and (ii) $b_{0} = 1.2$ (top-right panel). The bottom panels are the corresponding differences between the ACSs of the BHs, i.e., $\sigma_{\rm{dif}} \equiv \sigma_{\rm{LQG}} - \sigma_{\rm{RN}}$. To facilitate comparisons, in the bottom panels, we also have included an inset. Moreover, notice that the case $b_{0} = 1$ corresponds to the RN BH geometry.}
    \label{SACS}
\end{centering}
\end{figure*}

In Fig.~\ref{SACS}, we exhibit the situation where we have LQG and RN BHs with similar total ACSs. Notice that for the same choices of $Q/M$, the low- and high-frequency regimes will be very similar since the spacetimes share the same horizon and shadow radius structure. The main differences occur in the intermediate frequency regime, where we notice that the increase of $b_{0}$ typically increases the total ACSs; see Fig.~\ref{tacsdifb}. Consequently, higher values of $b_{0}$ contribute to increasingly different ACSs in the intermediate frequency regime.

\section{Final Remarks}\label{sec:remarks}

LQG-inspired BHs can be seen as promising scenarios for testing signatures of LQG in BH physics. Here, we have investigated in detail the absorption properties of massless test scalar waves in the background of a charged LQG BH, combining analytical and numerical methods. We sought to understand the role played by the quantum parameter $b_{0}$ and the BH charge-to-mass ratio $Q/M$ in the absorption spectrum. For completeness, we also computed the GBFs and reviewed the derivation and main properties of the charged LQG BH.

The main results of this work can be summarized as:

\begin{itemize}

\item The bound method used in Sec.~\ref{subsec:acsgbf} to obtain a semi-analytical bound for the GBFs is useful for dealing with the problem analytically. For example, it correctly predicts the behavior of the absorption probability as we vary $b_{0}$, considering $l = 0$ (cf. Fig.~\ref{gbf0}). However, the method is limited, as also pointed out in Ref.~\cite{Nakarachinda:2025hcd};

\item The total ACS oscillates around the GCS, and the sinc approximation reproduces the oscillatory pattern of the total ACS for intermediate-to-high frequency ranges. This comparison works as a consistency check of our numerical results and validates their accuracy;

\item The numerical calculation of the GBFs, i.e., the absorption probability, shows that for $l \geq 1$, considering higher values of $b_{0}$ can enhance or diminish the GBFs, depending on the frequency range. This is in stark contrast to the $l = 0$ case, which increases monotonically with frequency. Moreover, this effect does not occur when we fix $b_{0}$ and vary $Q/M$;

\item We also observe that the monopole mode $l = 0$ provides the main contributions to the total ACS in the low-frequency regime. Furthermore, the total ACS increases from the low-frequency regime to the moderate-frequency regime as we consider higher values of $b_{0}$, for a fixed ratio $Q/M$, but tends to the same values in the low- and high-frequency regimes. In turn, the total ACS decreases as we consider higher values of $Q/M$, for a fixed $b_{0}$. Both behaviors are consistent with the analysis of the BH structure and the effective potential;

\item Regarding the influence of the quantum parameter on the cross section, we also observed that as we increase the parameter $b_{0}$, the peaks and troughs of the total ACS exhibit opposite behaviors, i.e., the curve related to the highest peak corresponds to the deepest troughs; 

\item We also investigated the so-called mimic setups. Our results indicate that the low- and high-frequency regimes of the total ACSs of the LQG and RN BHs can be very similar. However, the cross sections inherit a difference in the intermediate-frequency range, which increases as we consider higher values of $b_0$.

\end{itemize}

Our main findings can be seen as a first step toward a better understanding of the absorption properties (in particular, spectrum and absorption probability) of LQG-inspired BHs. Remarkably, we have seen that LQG BHs can lead to absorption and GBF patterns different from those observed for standard BHs in GR, considering (neutral, massless) scalar fields.

Due to the novelty of our work, several avenues for further investigation are possible. Firstly, we can investigate the scattering cross section, seeking to better understand the role of BH parameters in the scattering spectrum. Secondly, we have considered massless scalar waves. From this perspective, a natural next step is to consider charged and massive scalar fields. Then, we can investigate the absorption and scattering properties within this context. In this case, we have superradiance~\cite{Bekenstein:1973mi,Benone:2015bst,dePaula:2024xnd}, i.e., the wave can be scattered with more energy than it originally had due to the electromagnetic interaction between the BH and the charged scalar wave.

Thirdly, in an astrophysical environment, BHs are typically surrounded by electromagnetic and gravitational fields. Therefore, it would be interesting to investigate the absorption and scattering cross sections in these scenarios. Finally, we also point out that we have considered a static and spherically symmetric setup. Generalizations of the results obtained here for rotating LQG-inspired BHs are also interesting since astrophysical BHs are expected to be spinning. Notice also that the behavior of the GBFs of the LQG BH spacetime may be of interest for other observables that require the determination of GBFs, such as the particle emission rate~\cite{ashraf2025thermal,araujo2026rotating}.

\begin{acknowledgments}

\end{acknowledgments}

The authors thank Luís Carlos Bassalo Crispino for useful comments and discussions. We acknowledge Conselho Nacional de Desenvolvimento Cient\'ifico e Tecnol\'ogico (CNPq), from Brazil, for partial financial support. V. B. Bezerra is partially supported by CNPq through the Research Project No. 307211/2020-7. M. A. A. de Paula is supported by CNPq/PDJ 150589/2025-5.

\appendix

\bibliography{ref}

\end{document}